\documentclass[10pt]{iopart}

\usepackage{iopams} 

\expandafter\let\csname equation*\endcsname\relax
\expandafter\let\csname endequation*\endcsname\relax
\usepackage{amsmath}

\usepackage[T1]{fontenc}
\usepackage[utf8]{inputenc} 
\usepackage{graphicx}
\usepackage[]{hyperref}
\usepackage[version=3]{mhchem}
\usepackage{subcaption}
\usepackage{amssymb}
\usepackage{gensymb}
\usepackage{longtable}
\usepackage{multirow}
\usepackage{mathtools}
\usepackage{booktabs}

\usepackage{upgreek}

\usepackage{epstopdf}
\usepackage{caption}

\usepackage{color}
\usepackage{xcolor}

\usepackage{graphicx}
\usepackage[labelformat=simple]{subcaption} 

\usepackage{placeins}
\usepackage[font={small}]{caption}

\usepackage{mathrsfs}

\usepackage{enumitem}
\usepackage{url}
\usepackage{times}

\usepackage{pgfplots}
\usepgfplotslibrary{groupplots}
\usepackage{tikz}
\usetikzlibrary{patterns}
\usetikzlibrary{external}
\usetikzlibrary{spy}

\usepackage{pgfkeys}
\usetikzlibrary{intersections}
\usepackage{listings}

\newlength\Colsep
\newcommand{\ex}{\hat{\vec e}_{x}}
\newcommand{\ey}{\hat{\vec e}_{y}}
\newcommand{\ez}{\hat{\vec e}_{z}}

\renewcommand{\vec}[1]{\boldsymbol{#1}}

\newcommand{\paren}[1]{\left(#1\right)}
\newcommand{\parenangle}[1]{\left\langle#1\right\rangle}

\newcommand{\volInt}[3]{\paren{#1 \, , #2}_{#3}}

\newcommand{\surInt}[3]{\parenangle{#1 \, , #2}_{#3}}

\newcommand{\grad}{\textrm{\textbf{grad}}\, }

\renewcommand{\div}{\textrm{div}\, }

\newcommand{\curl}{\textrm{\textbf{curl}}\, }
\newcommand{\curlOnly}{\textrm{\textbf{curl}}}

\renewcommand{\b}{\vec b}

\newcommand{\n}{\vec n}
\newcommand{\h}{\vec h}

\renewcommand{\e}{\vec e}

\renewcommand{\j}{\vec j}

\newcommand{\dt}{\partial_t}
\newcommand{\ec}{e_{\text{c}}}
\newcommand{\jc}{j_{\text{c}}}

\newcommand{\hs}{\vec h_{\text{s}}}

\newcommand{\jcoupling}{\vec j_{\text{cc}}}
\newcommand{\hcoupling}{\vec h_{\text{cc}}}
\newcommand{\hd}{\vec h_{\text{d}}}

\newcommand{\jcc}{\j_{\text{cc}}}
\newcommand{\ecc}{\e_{\text{cc}}}
\newcommand{\jccdyn}{\j_{\text{cc,d}}}

\renewcommand{\O}{\Omega}

\newcommand{\Oc}{\Omega_{\text{c}}}
\newcommand{\Occ}{\Omega_{\text{c}}^{\text{C}}}
\newcommand{\Om}{\Omega_{\text{m}}}
\newcommand{\Of}{\Omega_{\text{f}}}
\newcommand{\Och}{\Omega_{\text{ch}}}
\newcommand{\Ofi}{\Omega_{\text{f},i}}

\newcommand{\Gout}{\Gamma_{\text{out}}}

\newcommand{\hsppe}{\mathcal{H}_{\perp}}

\newcommand{\hspzpe}{\mathcal{H}_{\perp, 0}}

\newcommand{\hpf}{$h$-$\phi$-formulation\ }

\newcommand{\hpfOnly}{$h$-$\phi$-formulation}

\newcommand{\hsp}{\mathcal{H}}

\newcommand{\vsp}{\mathcal{U}}

\newcommand{\hspz}{\mathcal{H}_{0}}

\newcommand{\vspz}{\mathcal{U}_{0}}

\newcommand{\hspd}{\mathcal{H}^{\delta}}

\newcommand{\vspd}{\mathcal{U}^{\delta}}

\newcommand{\nodes}{\mathcal{N}}
\newcommand{\edges}{\mathcal{E}}

\usepackage{eso-pic}

\newcommand{\Nf}{N_{\text{f}}}

\newcommand{\method}{CATI\ }

\newcommand{\axial}{AI\ }
\newcommand{\transverse}{TI\ }

\usepackage{color}
\usepackage{xcolor}
\definecolor{myred}{rgb}{0.7,0.15,0.15}
\definecolor{mymaincolor}{rgb}{0.24, 0.36, 0.64}
\definecolor{mysecondcolor}{rgb}{0.21, 0.64, 0.87}
\definecolor{myblue}{rgb}{.2,0.45,0.5} 
\definecolor{myorange}{rgb}{0.78,0.6,0.3}
\definecolor{mygreen}{rgb}{.2,0.38,0.16}
\definecolor{myalert}{rgb}{0.97,0.09,0.21}
\definecolor{myformulation}{rgb}{0.33, 0.29, 0.31}
\definecolor{myformulation_back}{rgb}{1, 0.97, 0.91}
\definecolor{hf}{rgb}{0.93, 0.57, 0.13} 
\definecolor{hf_2}{rgb}{1.0, 0.89, 0.77} 
\definecolor{hf_3}{rgb}{1.0, 0.22, 0.0} 
\definecolor{hf_4}{rgb}{1.0, 0.4, 0.1} 
\definecolor{burlywood}{rgb}{0.87, 0.72, 0.53}
\definecolor{burntorange}{rgb}{0.8, 0.33, 0.0}
\definecolor{burntsienna}{rgb}{0.91, 0.45, 0.32}
\definecolor{af}{rgb}{0.4, 0.53, 0.34}
\definecolor{af_2}{rgb}{0.74, 0.77, 0.47}
\definecolor{af_3}{rgb}{0.12, 0.3, 0.17}
\definecolor{af_4}{rgb}{0.03, 0.34, 0.25}
\definecolor{haf}{rgb}{0.6, 0.51, 0.48}
\definecolor{haf_2}{rgb}{1, 0.97, 0.91}
\definecolor{taf}{rgb}{0, 0.55, 0.5}
\definecolor{ajf}{rgb}{0.29, 0.59, 0.82}
\definecolor{hbf}{rgb}{0.87, 0.36, 0.51}
\definecolor{prussianblue}{rgb}{0.0, 0.19, 0.33}
\definecolor{regalia}{rgb}{0.32, 0.18, 0.5}
	
\definecolor{myred}{rgb}{0.7,0.15,0.15}
\definecolor{mygreen}{rgb}{0.13,0.55,0.13}
\definecolor{myblue}{rgb}{0.25,0.41,0.88}

\definecolor{vir_0}{rgb}{0.993248, 0.906157, 0.143936}
\definecolor{vir_1}{rgb}{0.565498, 0.84243 , 0.262877}
\definecolor{vir_2}{rgb}{0.20803 , 0.718701, 0.472873}
\definecolor{vir_3}{rgb}{0.128729, 0.563265, 0.551229}
\definecolor{vir_4}{rgb}{0.190631, 0.407061, 0.556089}
\definecolor{vir_5}{rgb}{0.267968, 0.223549, 0.512008}
\definecolor{vir_6}{rgb}{0.267004, 0.004874, 0.329415}

\begin{document}

\title{Coupled Axial and Transverse Currents Method for Finite Element Modelling of Periodic Superconductors}

\author{Julien~Dular$^\text{1}$, Fredrik~Magnus$^\text{1,2}$, Erik~Schnaubelt$^\text{1,3}$, Arjan~Verweij$^\text{1}$, Mariusz~Wozniak$^\text{1}$
}
\address{$~^\text{1}$ CERN, Geneva, Switzerland\\
$~^\text{2}$ Norwegian University of Science and Technology, Trondheim, Norway\\
$~^\text{3}$ Technical University of Darmstadt, Darmstadt, Germany}
\ead{julien.dular@cern.ch}
\vspace{10pt}
\begin{indented}
\item[]\today
\end{indented}

\begin{abstract}
In this paper, we propose the Coupled Axial and Transverse currents (I) (CATI) method, as an efficient and accurate finite element approach for modelling the electric and magnetic behavior of periodic composite superconducting conductors. The method consists of a pair of two-dimensional models coupled via circuit equations to account for the conductor geometrical periodicity. This allows to capture three-dimensional effects with two-dimensional models and leads to a significant reduction in computational time compared to conventional three-dimensional models. After presenting the method in detail, we verify it by comparison with reference finite element models, focussing on its application to twisted multifilamentary superconducting strands. In particular, we show that the \method method captures the transition from uncoupled to coupled filaments, with accurate calculation of the interfilament coupling time constant. We then illustrate the capabilities of the method by generating detailed loss maps and magnetization curves of given strand types for a range of external transverse magnetic field excitations, with and without transport current.
\end{abstract}
%




\vspace{2pc}
\noindent{\it Keywords: Reduced Order Method, LTS, Finite Element Method, AC losses.}


\ioptwocol


\section{Introduction}
\addcontentsline{toc}{section}{Introduction}

Today, almost 40 years after the discovery of high-temperature superconductors (HTS), low-temperature superconductors (LTS) still completely dominate the market of superconducting applications in terms of the installed base for conductor manufacturing and the total number of magnets produced. For many applications, the LTS option has a lower cost when combined capital and operating costs are considered~\cite{teyber2020thermoeconomic}.

Although the dominance of LTS continues, the justification for using them is under increasing pressure due to the rising cost of helium. The LTS magnet industry is transitioning to use cryogenic configurations with reduced helium amount to offset its rising cost~\cite{lvovsky2013novel}. A common feature of these new cooling configurations is a reduced ability to absorb transient power loss, as there is no longer a large amount of helium able to provide the thermal buffer during periods of increased power loss, for example, during magnet ramp-up or down.

Besides, there is a continuous effort to develop high-field magnets based on LTS, particularly Nb$_3$Sn conductors and accelerator magnets~\cite{hfm_url}. Such magnets typically have an increased stored magnetic energy density, which in turn requires higher performance quench protection methods to ensure safe operation during sudden loss of superconducting state. As the magnetic field and stored energy density increase, the quench protection transients become generally shorter with much higher current and field change rates. In addition, there is an increased interest in and application of quench protection methods that rely on inducing a high current and/or field change rate in a magnet to quickly transition a large part of it to a normal state~\cite{ravaioli2015cliq, mulder2023external, ravaioli2023optimizing}. 

This ongoing change in cryogenic systems and quench protection methods brings renewed interest in calculating the power loss (AC loss) of LTS conductors, usually composite superconducting strands made of a large number of twisted superconducting filaments, or filament bundles, embedded in a conducting matrix. Their modelling is already served by a spectrum of methods.

There are long-established analytical methods that rely on simplifying assumptions~\cite{carr1974ac, wilson1983superconducting, morgan1970theoretical, turck1979coupling, ogasawara1980transient, campbell1982general} and have in general low computational cost. It comes, however, at the price of neglecting many details about the conductor, for example, the exact shape of the conductor or its filaments, the presence, location and thickness of diffusion barriers, the spatial distribution of the purity (RRR) of the copper stabilizer, and the local variation of the critical current of the filaments. To some degree, some of the analytical methods account for these features. However, if this is combined with a more complex excitation, for example, simultaneous change of transport current and transverse magnetic field, the analytical methods quickly reach their limits and, in practice, result in lower accuracy. The achievable accuracy may no longer be adequate for magnets that operate with a reduced amount of helium and/or rely on a quench protection method dominated by and designed with reliance on a specific magnitude of AC loss. As a result, methods accounting for more conductor details and being more accurate are needed.

These detailed models rely on the three-dimensional (3D) finite element (FE) method~\cite{zhao20173d, Grilli2014}. The computational power of modern computers enables such an approach, but model preparation and solution times remain challenging in practical use~\cite{escamez20163, riva2023h}. Parallelization methods are being considered to reduce the simulation time by sharing the work among numerous processing units~\cite{olm2019simulation, riva2023h, schnaubelt2024parallel}. However, the computational cost is still high, and methods based on 3D FE models are currently not practical for in-depth analysis of multiple scenarios, which often need to be considered for the optimization of the conductor as part of the magnet design.

Between these two ends of the spectrum of AC loss calculation are the so-called reduced order methods, which represent more details about the conductor than analytical methods, but with some special approach to reduce the computational cost compared to 3D FE models.

Reduced order models based on a helicoidal change of variables~\cite{nicolet2004modelling, stenvall2012current, dular2023helicoidal}, leading to special two-dimensional (2D) models, are one possibility, but they are limited to helicoidally symmetric conductor cross sections and remain challenging to implement in the case of transverse field excitation, i.e., perpendicular to the strand axis, with nonlinear materials~\cite{dular2023helicoidal}. Treating such excitations is however crucial in the context of AC loss calculation.

Other reduced order models include the use of a Frenet frame to simplify the geometry definition~\cite{kameni2019reduced}, or homogenization techniques with anisotropic material properties~\cite{zhao20173d}. Nonlinear circuit models were also proposed in~\cite{soldati2024ac}, with lumped circuit elements evaluated with preliminary FE resolutions~\cite{breschi2008electromagnetic}.

The method proposed by T. Satiramatekul and F. Bouillault in \cite{satiramatekul2010numerical, satiramatekul2007magnetization, satiramatekul2005contribution} reproduces the effect of coupling currents in the conducting matrix by introducing equivalent resistances between the filaments, accounting for the twist pitch length of the wire. This allows to reproduce uncoupled, partially coupled, and coupled filament regimes without using a 3D model, but a simple, yet appropriate, 2D approach. The method presented in this contribution is inspired by that work.

We propose to consider a pair of 2D models: one for calculating axial currents, flowing along the wire axis, and one for modelling coupling currents, or transverse currents, flowing perpendicular to the wire axis, in the matrix in-between the superconducting filaments. We then couple these models using circuit equations that represent the effect of the periodic geometry of the conductor. We refer to this approach as the \method method, standing for Coupled Axial and Transverse currents (I) method.

A key feature of the \method method is that it is not only applicable to helicoidally symmetric conductors, like round twisted composite conductors, but more generally to all conductors with a periodic geometrical structure. We show that it produces accurate results, with a drastic reduction of computational cost for performing the simulation compared to conventional 3D models.

Another advantage of the method is the relative ease of its implementation. It can be implemented in any FE software allowing for field-circuit coupling. In this work, the method is implemented in GetDP~\cite{getdp} within FiQuS~\cite{vitrano2023open}, developed at CERN as part of the STEAM framework~\cite{Bortot2017}. Geometry and mesh generation is performed by Gmsh~\cite{gmsh}. All the software is open-source and free to use\footnote{[Online] \url{https://cern.ch/fiqus}.}. The input files for the simulations are provided in \cite{cati_steamAnalysis} so that the \method results can be reproduced.

In Section~\ref{sec_lfm}, we present the \method method, describe in detail its practical implementation, and highlight its underlying assumptions and limitations. We then verify and assess the range of applicability of the approach in two steps. First, in Section~\ref{sec_verification}, we consider a linear problem with constant material properties, and discuss the accuracy of the method, focussing on the coupling current dynamics. Second, in Section~\ref{sec_verification_nonlinear}, we consider a nonlinear problem with realistic field-dependent material parameters and verify the prediction of the \method method against those of a 3D FE reference model. Finally, in Section~\ref{sec_applications}, we exploit the method and demonstrate its capabilities on a helicoidally symmetric strand and on a non-helicoidally symmetric, but periodic, wire-in-channel geometry.

\section{\method Method}\label{sec_lfm}

\begin{figure}[h!]
\begin{center}
\includegraphics[width=0.9\linewidth]{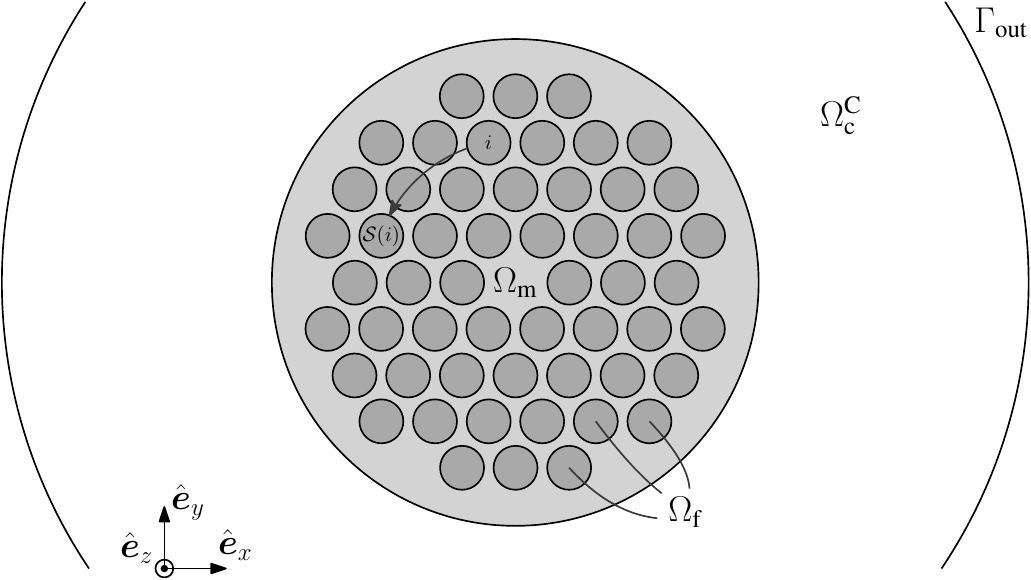}
\caption{Computational domain for the cross section of a superconducting strand ($\Nf=54$). The 3D geometry is generated by a rotated extrusion of the cross section along $\ez$ with twist pitch length $p$ (see Fig.~\ref{3Dgeo_54fil_mesh_full}). The curved arrow represents the rotation of filament $i$ to filament $\mathcal{S}(i)$ along the periodicity length $\ell = p/6$.}
\label{domain_definition}
\end{center}
\end{figure}

\begin{figure}[h!]
\begin{center}
\includegraphics[width=\linewidth]{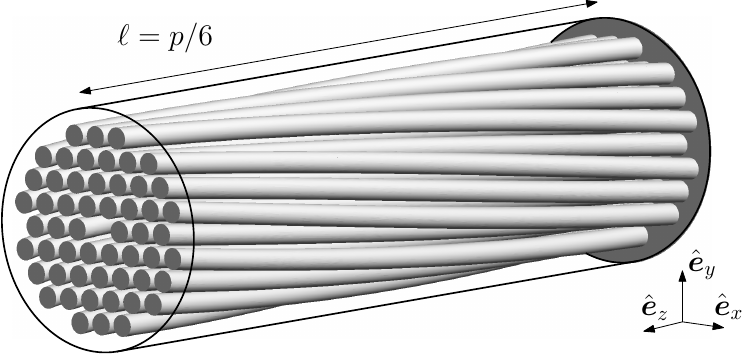}
\caption{3D representation of the 54-filament composite strand whose transverse cross section is represented in Fig.~\ref{domain_definition}. The cylindrical outline represents the boundary of the conducting matrix in which the filaments are embedded.}
\label{3Dgeo_54fil_mesh_full}
\end{center}
\end{figure}

We consider a 2D cross section of a periodic composite conductor, perpendicular to the conductor main axis $\ez$. The cross section contains $\Nf\in \mathbb{N}_0$ filaments $\Ofi$, with $i\in F = \{1,\dots,\Nf\}$. The filament positions are assumed to be periodically appearing in exactly the same locations after a given displacement along $\ez$. For a round strand such as the one shown in Figs.~\ref{domain_definition} and \ref{3Dgeo_54fil_mesh_full}, this is achieved after twisting the strand with a twist pitch length $p$.

The union of the filaments defines the domain $\Of = \cup_{i\in F} \Ofi$, embedded in a conducting matrix $\Om$. The conducting domain $\Oc = \Of\cup \Om$ is surrounded by a non-conducting domain $\Occ$, whose external boundary is $\Gout$.

The \method method is based on the assumption that there is a periodicity of the cross sections along $\ez$. Let $\ell$ be the periodicity length, that is, the distance along $\ez$ after which every filament takes the position of another one. Let the link between the positions in successive cross sections, separated by a distance $\ell$, be described by a permutation operator $\mathcal{S}$, defining a bijection of the set of filaments onto itself,
\begin{align}
\mathcal{S}\, :\, F\to F,\ i \to \mathcal{S}(i).
\end{align}
For example, with the geometry in Figs.~\ref{domain_definition} and \ref{3Dgeo_54fil_mesh_full}, the smallest possible value of $\ell$ is $p/6$, and filament $\Omega_{\text{f},\mathcal{S}(i)}$ is the image of filament $\Ofi$ under a $2\pi/6$ rotation around the strand center (see the dashed arrow in Fig.~\ref{domain_definition}).

The periodicity does not need to be created by twisting the geometry; the method only requires a periodic structure of the cross sections. Examples of such periodic, but not twisted, cross sections include common cable types, like Rutherford~\cite{verweij1997electrodynamics} or Roebel~\cite{goldacker2014roebel}. The general concept of the method applies to them. However, they are outside of the scope of this contribution.

In the following, vectors parallel to $\ez$ are referred to as axial vectors, whereas those that are perpendicular to $\ez$ are referred to as transverse vectors.

In Section~\ref{sec_weakForm}, we define the strong and weak forms of the method and discuss the assumptions that are made. In Sections~\ref{sec_correctedLength} and \ref{sec_dynamicCorrection} we propose two improvements of the method. Both are optional in the implementation, but both improve significantly the accuracy of the model. Finally, in Section~\ref{sec_discreteImplementation}, we focus on the practical implementation of the method, by describing in detail the spatial discretization of the unknown fields with finite elements.

\subsection{\method method equations}\label{sec_weakForm}

\begin{figure}[h!]
\begin{center}
\includegraphics[width=\linewidth]{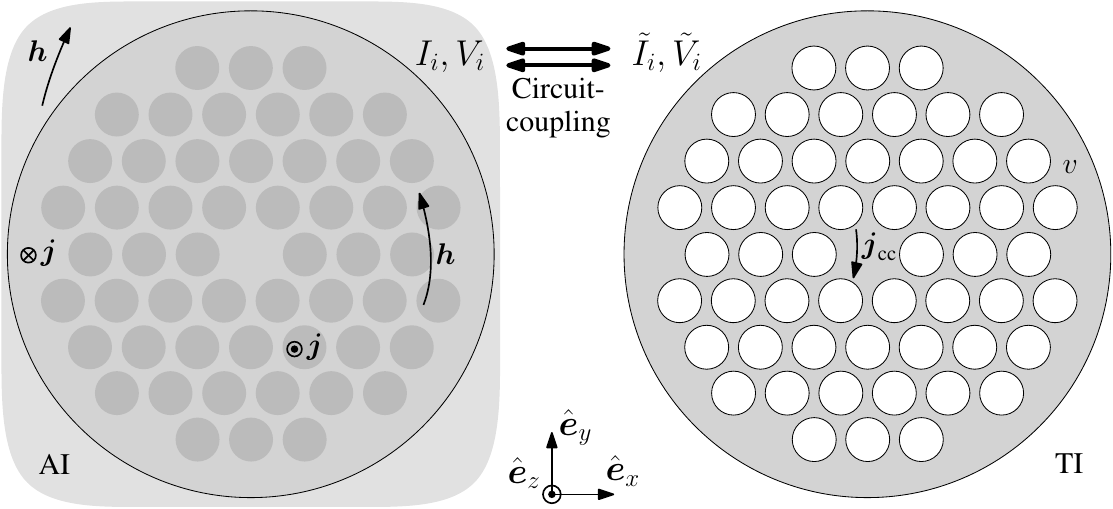}
\caption{Illustration of the \axial and \transverse models fields. (Left) \axial model: solves for axial currents $\j$ in $\Oc$ and transverse magnetic field $\h$ in $\O$. (Right) \transverse model: solves for transverse currents $\jcc$ in $\Om$ only. Global quantities (currents $I_i$ and $\tilde I_i$, and voltages $V_i$ and $\tilde V_i$) interface both models via circuit-coupling equations.}
\label{IP_OOP_conceptual}
\end{center}
\end{figure}

The \method method consists of a pair of 2D models: (i) a model describing axial currents (along $\ez$) in $\Oc = \Of \cup\Om$, and the associated transverse magnetic field (perpendicular to $\ez$) in $\O = \Oc \cup \Occ$, and (ii) a model describing transverse currents (perpendicular to $\ez$) in $\Om$ only. This is illustrated in Fig.~\ref{IP_OOP_conceptual}. In the following, we refer to them as the \axial and \transverse models, standing for axial currents and transverse currents models, respectively. Net currents and voltages are defined for each filament in both models, they are referred to as global quantities. The \axial and \transverse models are coupled via these global quantities using circuit equations that are defined in order to account for the periodicity of the geometry, hence of its three-dimensional aspect. In particular, the periodicity length $\ell$ is explicitly present in the governing equations.

In this section, we define the strong and weak forms of the problem in the continuous setting. We successively introduce the \axial model, the \transverse model, and the circuit equations. We conclude by discussing the assumptions behind the equations.

\subsubsection{Axial currents formulation}
The \axial model is governed by Maxwell's equations in the magnetodynamic (or magneto-quasistatic) regime~\cite{jackson1999classical}, whose strong form reads:
\begin{equation}\label{MQSequations}
\left\{\begin{aligned}
\div\b &= 0,\\
\curl\h &= \j,\\
\curl\e &= -\dt \b,
\end{aligned}\right. \quad \text{with} \quad  \left\{\begin{aligned}
\b &= \mu\, \h,\\
\e &= \rho\, \j,
\end{aligned}\right.
\end{equation}
with $\b$, $\h$, $\j$ $\e$, $\mu$, and $\rho$, the magnetic flux density (T), the magnetic field (A/m), the current density (A/m$^2$), the electric field (V/m), the permeability (H/m), and the resistivity ($\O$m), respectively. In the non-conducting domain, $\rho \to \infty$ and $\j = \vec 0$, and Ampère's law reads $\curl \h = \vec 0$.

We define $\j$ and $\e$ as axial vectors, and $\h$ and $\b$ as transverse vectors. The chosen weak form of the model is that of a classical 2D \hpfOnly, in order to ensure the numerical efficiency in the treatment of the superconductor power law for the resistivity~\cite{dular2019finite}. The 2D formulation is integrated along $\ez$ over a length $2\ell$, assuming that the solution is constant over this length.

Denoting by $\volInt{\vec f}{\vec g}{\O}$ the integral over $\O$ of the dot product of any two vector fields $\vec f$ and $\vec g$, the weak form reads: from an initial solution at $t=0$, find $\h\in\hsp(\O)$ such that, for $t>0$ and $\forall \h' \in \hspz(\O)$, we have~\cite{dular2019finite}
\begin{align}\label{eq_oop}
\volInt{2\ell\, \dt(\mu\, \h)}{\h'}{\O} + \volInt{2\ell\, \rho\, \curl \h}{\curl \h'}{\Oc}\qquad~\notag\\
 = V_\text{t} \mathcal{I}_{\text{t}}(\h') + \sum_{i\in F} V_i \mathcal{I}_i(\h').
\end{align}
The function space $\hsp(\O)$ is the subspace of $H(\curlOnly;\O)$ containing transverse vector functions that are curl-free in $\Occ$ and fulfill appropriate boundary conditions. This space, together with its link with global variables will be defined explicitly in the spatial discretization step. The space $\hspz(\O)$ is the same space as $\hsp(\O)$ but with homogeneous boundary conditions.

The operator $\mathcal{I}_i(\h)$ gives the circulation of $\h$ around filament $\Ofi$, which is the net current, denoted by $I_i$, flowing in that filament. The associated voltage is denoted by $V_i$. It has units volt (V) because of the multiplication by $2\ell$, which differs from the usual 2D \axial formulations where the voltage is in volt per unit length (V/m). The operator $\mathcal{I}_{\text{t}}(\h)$ is the net current flowing in the whole conductor, i.e., the transport current, denoted by $I_{\text{t}}$, which is the sum of the currents in all the filaments and in the conductor matrix. The associated voltage is denoted by $V_{\text{t}}$, it is the voltage applied by the power generator over a length $2\ell$. Currents and voltages $I_i$, $V_i$, $i\in F$, $I_\text{t}$, and $V_\text{t}$, are referred to as the \axial model global quantities.

In reality, the solution is not constant over the distance $2\ell$, but rather varies smoothly along the wire. We discuss how to correct for this assumption in Section~\ref{sec_correctedLength}. Also, integrating over $2\ell$ is a choice. Integrating over $\ell$ instead leads to an equally valid model and is discussed in Section~\ref{sec_couplingEquations}.

\subsubsection{Transverse currents formulation}
The \transverse model is governed by direct current flow (electrokinetics) equations~\cite{jackson1999classical} whose strong form reads:
\begin{equation}\label{EKequations}
\left\{\begin{aligned}
\div\jcc &= 0,\\
\curl\ecc &= \vec 0,
\end{aligned}\right. \quad \text{with} \quad \jcc = \sigma\, \ecc,
\end{equation}
with $\jcc$ the coupling current density (A/m$^2$), $\ecc$ the coupling current electric field (V/m), and $\sigma = \rho^{-1}$ the electric conductivity (S/m). The model is solved in $\Om$ only, and both $\jcc$ and $\ecc$ are defined as transverse vectors.

The \transverse model is written in terms of the electric scalar potential $v \in \vsp(\Om)$, with $\ecc = -\grad v$, so that the equation $\curl \ecc = \vec 0$ is satisfied by construction. The function space $\vsp(\Om)$ is a subspace of $H^1(\Om)$ containing functions that are region-wise constant on filament boundaries $\partial \Ofi$ (this is justified in Section~\ref{sec_assumptions}). We denote these voltage values as $v|_{\partial \Ofi} = \tilde{\mathcal V}_i(v) = \tilde V_i$, for $i \in F$. Together with their associated current, $\tilde I_i$, for $i\in F$, defined in the following, they constitute the \transverse model global quantities.

The \transverse model is a weak form of the divergence-free condition on the current density $\jcc$, with
\begin{align}\label{eq_jcc}
\jcc = -\sigma\, \grad v,
\end{align}
integrated along $\ez$ over a length $2\ell$, assuming that the solution is $z$-independent over this length. We write, in $\Om$, and $\forall v'\in \vspz(\Om)$, 
\begin{align}
&-\volInt{2\ell\ \div(\sigma\, \grad v)}{v'}{\Om} = 0,\\
\Leftrightarrow & \volInt{2\ell\, \sigma\, \grad v}{\grad v'}{\Om} + \surInt{2\ell\, \jcc\cdot\n}{v'}{\partial\Om} = 0,
\end{align}
with $\n$ the unit outer normal vector to the wire matrix surface and $\partial \Om$ the boundary of $\Om$, including $\partial \Ofi$, $\forall i\in F$, and the external matrix boundary. The second term of the left-hand side is zero on the external matrix boundary because $\jcc\cdot\n = 0$ on this surface. On the filament boundaries $\partial\Ofi$, $i\in F$, we define
\begin{align}
\tilde I_i = 2\ell\, \int_{\partial\Ofi} \jcc \cdot \n\, dS,
\end{align}
that is, $\tilde I_i$ is the net current entering filament $i$ over the length $2\ell$. Using the fact that $v'$ is constant on $\partial \Ofi$, the weak form can therefore be stated as follows: find $v \in \vsp(\Om)$ such that, $\forall v'\in \vspz(\Om)$, we have
\begin{align}\label{eq_ip}
\volInt{2\ell\, \sigma\, \grad v}{\grad v'}{\Om} + \sum_{i\in F} \tilde I_i \tilde{\mathcal V}_i(v') = 0.
\end{align}
Note that $\tilde I_i$ are currents in ampère (A) and not currents per unit length (A/m) due to the multiplication by $2\ell$ in the weak formulation.

\subsubsection{Circuit-coupling equations}\label{sec_couplingEquations}
The \axial and \transverse models involve $4\Nf+2$ global quantities. There are already $2\Nf+1$ equations linking them in Eqn.~\eqref{eq_oop} and \eqref{eq_ip}, so that only $2\Nf+1$ equations are still to be defined. One equation consists in setting directly either the transport current $I_\text{t}$ or the associated voltage $V_\text{t}$. The remaining $2\Nf$ equations are defined below; they couple the \axial and \transverse models and are the core of the \method method. They rely on the periodicity of the geometry of the conductor.

\begin{figure}[h!]
\begin{center}
\includegraphics[width=\linewidth]{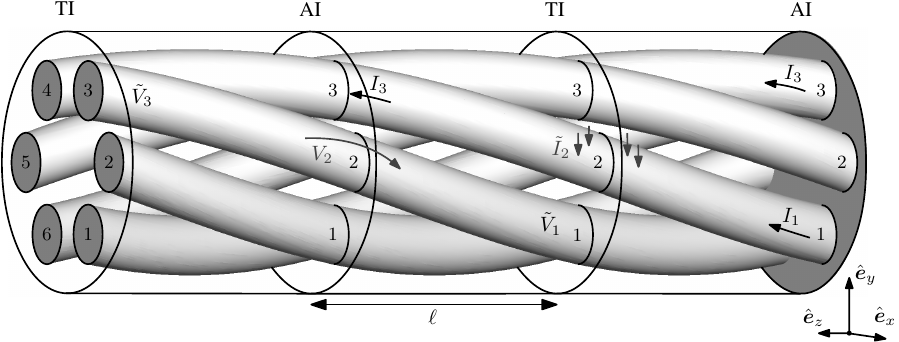}
\caption{Interpretation of Eqn.~\eqref{eq_coupling_I} and \eqref{eq_coupling_V} on a 6-filament periodic geometry, with $\ell = p/6$, $i=1$, $j = \mathcal{S}(i) = 2$, and $k = \mathcal{S}(\mathcal{S}(i)) = 3$.}
\label{global_eqn_illustration_6_filaments}
\end{center}
\end{figure}

Using the permutation operator $\mathcal S$, for every $i\in F$ we define $j = \mathcal{S}(i) \in F$, and $k = \mathcal{S}(j) = \mathcal{S}(\mathcal{S}(i))$, and we write:
\begin{align}
\tilde I_{j} &= I_k - I_i,\label{eq_coupling_I}\\
V_{j} &= \tilde V_{k} - \tilde V_i.\label{eq_coupling_V}
\end{align}
We illustrate the meaning of Eqn.~\eqref{eq_coupling_I} and \eqref{eq_coupling_V} on a simple example with $\Nf = 6$ filaments in Fig.~\ref{global_eqn_illustration_6_filaments}. Using the periodicity of the cross sections, the difference between the currents $I_3$ and $I_1$ is equal to the current $\tilde I_2$, accumulated over the length $2\ell$. Similarly, the difference between the voltages $\tilde V_3$ and $\tilde V_1$ is equal to the voltage $V_2$, accumulated over the length $2\ell$. The \axial and \transverse models can therefore be interpreted as being solved on staggered cross sections. By periodicity, all these cross sections are equivalent, which allows us to consider only one of them.

Note that it is possible to reduce the integration along $z$ from $2\ell$ down to $\ell$, so that the \axial and \transverse models are solved on distinct cross sections. In the case of Fig.~\ref{global_eqn_illustration_6_filaments}, this leads to \transverse cross sections that are rotated by an angle of $\pi/6$ with respect to the \axial ones. For the geometries considered in this contribution, we observed that the obtained results (with adapted corrected length, as discussed in Section~\ref{sec_correctedLength}) are almost identical.


\subsubsection{Underlying assumptions}\label{sec_assumptions}

The \method method requires a periodic geometry of the conductor, whose cross section repeats itself after a given length in the main direction of the conductor $\ez$. Along this length, filaments must move from one location to another, and hence be tilted with respect to $\ez$. Because purely 2D models are used to describe the fields, one main assumption of the \method method is that the effect of the tilt angle on the field distribution can be neglected.

With twisted conductors, this is valid provided that the twist pitch length $p$ is sufficiently large with respect to the radius of the outermost filament layer. This will be discussed in Section~\ref{sec_verification}. In particular, it will be shown that for realistic twist pitch lengths, this approach is justified.

The \method method also assumes that the filament resistivity is sufficiently small compared to the matrix resistivity for considering that the \transverse model voltage $v$ is constant on each filament boundary. However, if this is not the case, axial currents will be more evenly shared among the matrix and the filaments, and the contribution of coupling currents to total losses is therefore expected to decrease. In the extreme case of a quench, when the filaments become relatively highly resistive, most of the current flows in the matrix, and coupling currents become negligible. These situations can be handled by the \method method, as it describes the axial current flow in the matrix.

The heart of the method lies in the circuit-coupling equations. They describe the evolution of fields and global quantities in the filaments along $\ez$. This evolution is represented by the global quantities of distinct filaments within a single cross section, exploiting the periodicity of it. For the example in Fig.~\ref{global_eqn_illustration_6_filaments}, the evolution of global quantities of filaments can be decomposed in up to $6$ steps before a complete rotation is performed. A minimum of 2 steps is necessary for the method to be relevant.

Finally, the method presented here does not allow for modelling excitation with an external axial magnetic field. Further extensions are necessary to include axial external field effects in the method.

\subsection{Length correction factor}\label{sec_correctedLength}

The connections between twisted filaments are fully implemented by the circuit coupling equations. The twist pitch length appears in the weak formulations of the \axial and \transverse models, via the multiplication by $2\ell$ in Eqn.~\eqref{eq_oop} and \eqref{eq_ip}.

The use of 2D models leads to assuming that the filaments carry constant currents $I_i$ and maintain constant voltages $\tilde V_i$ over a distance $2\ell$ along the wire, in the \axial and \transverse models, respectively, after which the position of one filament suddenly changes to the next one, and so do the values of the currents and voltages. Similarly, the voltages $V_i$ and the currents $\tilde I_i$ are obtained by simple multiplication of 2D solutions by $2\ell$.

\begin{figure}[h!]
\begin{center}
\includegraphics[width=\linewidth]{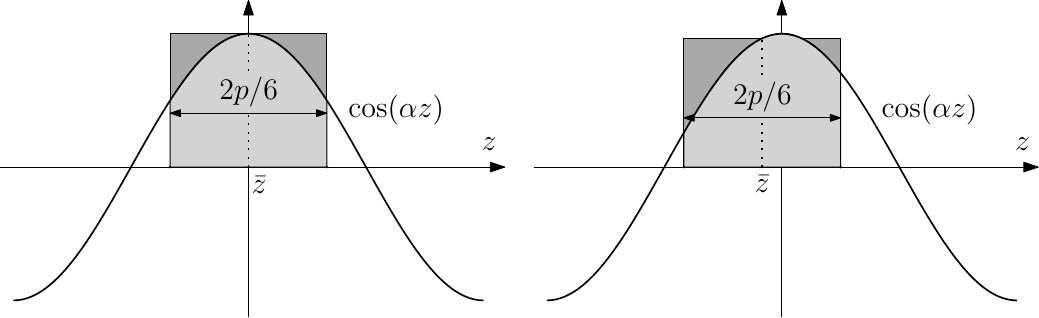}
\caption{Illustration of the overestimation factor due to the modelling of the \axial and \transverse models as pure 2D models, in the case $2\ell = 2p/6$, for two different positions $\bar z$. The rectangles of width $2\ell$ and height $\cos (\alpha\bar z)$, with $\alpha = 2\pi/p$, represent the integrated global quantities in the \method method, modelled as piecewise constant functions. The light-gray areas below the curve represent the actual integrated global quantities for round-shape twisted conductors, assuming a sinusoidal evolution.}
\label{correction_factor}
\end{center}
\end{figure}

In reality however, the variation of the global quantities is smooth and continuous. In round-shape conductors, these quantities are expected to vary in a sinusoidal manner along $\ez$~\cite{ries1977ac}. As illustrated in Fig.~\ref{correction_factor}, a pure 2D approach therefore leads to an overestimation of the integrated quantities.

To account for this effect, a correction factor can be evaluated analytically. With $\alpha = 2\pi/p$, at a given position $\bar z$, we have
\begin{align}\label{eq_corr_factor_dev}
\frac{\int_{\bar z-\ell}^{\bar z+\ell}\cos(\alpha z)\, dz}{\int_{\bar z-\ell}^{\bar z+\ell}\cos(\alpha \bar z)\, dz} &= \frac{\sin(\alpha (\bar z+\ell)) - \sin(\alpha(\bar z- \ell))}{2\alpha \ell \cos(\alpha \bar z)}\notag\\
&= \frac{2\cos(\alpha \bar z)\sin(\alpha\ell)}{2\alpha \ell \cos(\alpha \bar z)}\notag\\
&= \frac{\sin(\alpha\ell)}{\alpha \ell} = \frac{\sin(2\pi \ell/p)}{2\pi \ell /p},
\end{align}
where integrands are symmetrically integrated around $\bar z$ due to the symmetry of the circuit coupling equations. The fact that Eq.~\eqref{eq_corr_factor_dev} does not depend on $\bar z$ is convenient, as this implies that the overestimation of the twist effect is identical for all filaments, and can then in principle be corrected by a single correction factor. For many conductors used in practice, the filament structure is hexagonal, so that $\ell = p/6$ and the correction factor is equal to $\approx 0.8270$.

To implement this correction in the model, the periodicity length $\ell$ in Eqn.~\eqref{eq_oop} and \eqref{eq_ip} can be replaced by a reduced, or corrected, length $\ell^{\text{c}}$, whose value is given by
\begin{align}\label{eq_corr_factor}
\ell^{\text{c}} = \frac{\sin(2\pi \ell/p)}{2\pi \ell /p}\ \ell.
\end{align}
The effect of the modification is to directly rescale the global quantities. We will show in Section~\ref{sec_verification} that this correction strongly improves the accuracy of the model, helping to better reproduce the analytical time constant associated with the coupling currents.

In the alternative case where the equations are integrated over length $\ell$ instead of $2\ell$, leading to the \axial and \transverse models to be solved on rotated distinct cross sections, $2\ell$ must be replaced by $\ell$ in Eq.~\eqref{eq_corr_factor_dev}. In the common case in which $\ell = p/6$, this leads to a correction factor of $\approx 0.9549$. 

A similar approach can be followed for other conductor types, for which quantities are expected to follow a non-sinusoidal evolution along the conductor. It is however not guaranteed in such cases that a common correction factor can be derived for the whole cross section, as is the case above with the invariance of Eq.~\eqref{eq_corr_factor_dev} with respect to $\bar z$. An average correction factor may be considered in such cases.

\subsection{Coupling current dynamics}\label{sec_dynamicCorrection}

The transverse coupling current flow is described by direct current flow equations. These equations do not describe eddy currents, and the transverse current flow is therefore static (ohmic). This is a good approximation at low frequencies. However, at higher frequencies, coupling currents can no longer be described solely by an electrostatic potential, because of the inevitable appearance of non-negligible transverse eddy currents. This is illustrated by reference solutions in Fig.~\ref{couplingCurrent_dynamics_all_four}(b) in Section~\ref{sec_verification}, where coupling currents at a frequency of $1$ kHz are localized on the outer part of the matrix. Such a current flow clearly cannot be described by the gradient of a scalar potential, as in Eq.~\eqref{eq_jcc}. If nothing is done to correct for this, an error on the coupling current loss contribution is expected to increase as the frequency increases, as shown later in Section~\ref{sec_dynCorrEffect}.

Note that this effect is different from the classical skin effect in the matrix, associated with axial currents (for transverse external field considered here), which is already properly accounted for by the \axial model. The axial currents flowing close to conductor surface due to the skin effect constitute the dominant loss contribution at high frequencies, so that the above-mentioned error on coupling current dynamics is expected to be detrimental on a limited frequency range only, as it is shown later.

As explained below, a rigorous treatment of a dynamic coupling current flow using magnetodynamics equations can however only be done with a full 3D approach.

Let $\hcoupling$ be the magnetic field associated with the transverse coupling current density $\jcoupling$ in the matrix $\Om$ via the relation $\curl \hcoupling = \jcoupling$. Such a magnetic field is not purely axial. Indeed, a purely axial field cannot generate net transverse currents exiting or entering a filament. The field $\hcoupling$ must therefore also have $z$-dependent transverse components. Coupling currents are indeed a result of the twist, and are 3D effects by nature. As a result, while we will show that coupling currents, and particularly their power dissipation, can be well approximated using the 2D direct current flow equations and circuit-coupling equations, the associated magnetic field $\hcoupling$ cannot be retrieved using 2D models only. For a rigorous treatment of dynamic effects in the \transverse model, such a field is however necessary, as it appears in Faraday's law.

In order to evaluate the importance of dynamic effects in the coupling current distribution (and the resulting power loss), we therefore only propose an approximate correction method. It corrects the solution a posteriori and does not interfere with the resolution of the \axial and \transverse models. It can be seen as an optional post-processing correction, giving an estimation of the dynamic effect contribution.

The method consists in projecting the static transverse coupling current distribution $\jcc = -\sigma\, \grad v$ on the curl of an axial field function space~\cite{geuzaine1999galerkin}. To this end, we define a static magnetic field $\hs \in \hsppe(\Om)$ that satisfies, $\forall \h'_{\textrm{s}} \in \hsppe(\Om)$, the following weak form:
\begin{align}\label{eq_dynCorr_hs}
\volInt{\sigma\, \grad v}{\curl \h'_{\textrm{s}}}{\Om} + \volInt{\curl \hs}{\curl \h'_{\textrm{s}}}{\Om} = 0,
\end{align}
with $\hsppe(\Om)$ the subspace of $H(\curlOnly;\Om)$ containing axial vector fields (parallel to $\ez$). In practice, we gauge the function space by imposing a zero average value of $\hs$ over $\Om$, via a Lagrange multiplier~\cite{babuvska2003mixed}. Note that, locally, $\curl \hs \neq - \sigma\, \grad v$. Indeed, as mentioned above, an axial field alone cannot reproduce a transverse coupling current distribution with net current exchanges between the filaments. The field $\hs$ is only an approximation of $\hcoupling$ that belongs to $\hsppe(\Om)$.

Next, we solve for a dynamic magnetic field correction $\h_{\textrm{d}}\in \hspzpe(\Om)$ of this static magnetic field that satisfies a weak form of Faraday's law, i.e., $\forall \h_{\textrm{d}}'\in \hspzpe(\Om)$,
\begin{align}\label{eq_dynCorr_hd}
&\volInt{\dt(\mu(\hs + \h_{\textrm{d}}))}{\h'_{\textrm{d}}}{\Om} + \volInt{\rho\, \curl \h_{\textrm{d}}}{\curl \h'_{\textrm{d}}}{\Om} = 0.
\end{align}
The function space $\hspzpe(\Om)$ is the subspace of $\hsppe(\Om)$ of functions that vanish on the external matrix boundary, so that $\curl \h_{\textrm{d}}\cdot \n = 0$ on this boundary and $\h_{\textrm{d}}$ only induces circulations of currents inside the wire. Note the absence of $\hs$ in the curl-curl integral of Eq.~\eqref{eq_dynCorr_hd}, because $\hs$ is associated with a static current flow so that the associated electric field $\rho\,\curl\hs$ should be curl-free.

Finally, we replace the static coupling current density $\jcc$, from Eq.~\eqref{eq_jcc}, by the following (dynamic) coupling current density,
\begin{align}
\jccdyn = -\sigma\, \grad v + \curl \h_{\textrm{d}}.
\end{align}
The second term is the dynamic correction. It lets the global currents $\tilde I_i$ unchanged by construction (an axial magnetic field cannot generate a net current entering or exiting a filament). This correction produces more realistic coupling current distributions and improves the accuracy of power loss evaluation at frequencies of the order of $1$ kHz. This is illustrated and discussed in Section~\ref{sec_verification}.


\subsection{Discretization and implementation details}\label{sec_discreteImplementation}

For a numerical resolution, the \axial and \transverse formulations are solved on a discretized representation of the geometry, called a mesh. The continuous function spaces $\hsp(\O)$ and $\vsp(\Om)$, introduced in formulations Eqn.~\eqref{eq_oop} and \eqref{eq_ip}, are replaced by discrete functions spaces $\hspd(\O)\subset \hsp(\O)$ and $\vspd(\Om)\subset \vsp(\Om)$, whose generating functions are associated with elementary entities of the mesh (nodes, edges, or groups of them), using Whitney shape functions~\cite{bossavit1988whitney}. We denote by $\nodes(\O_i)$ and $\edges(\O_i)$, the set of nodes and edges, respectively, of the mesh in a given domain $\O_i$, including entities on the boundary of $\O_i$.

\subsubsection{Axial currents function space}

The magnetic field $\h$ of the \axial model is discretized as in a usual \hpf~\cite{dular2019finite}, but with a special treatment at the filament boundaries to extract the filament currents $I_i$ directly: the filament boundaries $\partial \Ofi$ are put formally in the non-conducting domain $\Occ$. Hence, the domain $\Occ$ contains the exterior of the strand and the union of the filament boundaries.

Then, as usual, edge functions $\vec w_e$ are used in $\Oc\backslash \partial \Oc$, gradients of node functions $w_n$ are introduced in $\Occ$, and cut functions~\cite{pellikka2013homology} are defined for the net currents. There is one cut function per filament, $\vec c_i$, and one cut function for the whole strand, $\vec c_{\text{t}}$. The magnetic field is therefore expressed as the following linear combination:
\begin{align}\label{eq_OOP_functionSpace}
\h = \sum_{e \in \edges(\Oc\backslash \partial \Oc)} h_{e} \ \vec w_e + \sum_{n \in \nodes(\Occ)}  \phi_n \ \grad w_n\quad&\notag\\
 + I_{\text{t}} \ \vec c_{\text{t}} + \sum_{i \in F} I_i \ \vec c_i,&
\end{align}
where coefficients $h_e$, $\phi_n$, $I_{\text{t}}$, and $I_i$ are the degrees of freedom (DOFs) defining the magnetic field $\h$ in the discrete function space $\hspd(\O)$. As a scalar field $\phi$ is introduced on filament boundaries that are not connected to the external $\Occ$ domain, the field $\phi$ should be gauged on each filament boundary, e.g., by setting it to zero at an arbitrary point on each $\partial \Ofi$. It also has to be gauged in the external $\Occ$ domain.

Note that Eq.~\eqref{eq_OOP_functionSpace} leads to an identical function space (just with a different basis) to that of a usual \hpf in which the filament boundaries are kept in the conducting domain $\Oc$. In the usual case however, there are no explicit cut functions $\vec c_i$ for each $i\in F$ and the filament currents can only be extracted via a combination of several DOFs, i.e., via the signed sum of coefficients $h_e$ associated with oriented edges $e \in \edges(\partial \Ofi)$. This leads to a less convenient implementation in GetDP, but may be considered as an alternative in other software.

\subsubsection{Transverse currents function space}

The electric scalar potential $v$ is discretized with node functions $w_n$ as follows:
\begin{align}\label{eq_IP_functionSpace}
v = \sum_{n \in \nodes(\Om\backslash \partial \Of)} v_n \ w_n + \sum_{i\in F} \tilde V_i \ q_i,
\end{align}
where the global shape function $q_i$ for $i\in F$ is defined as the sum of node functions associated with the nodes on filament boundary $\partial \Ofi$. This grouping of shape functions ensures that $v$ is constant on each filament boundary $\partial \Ofi$, with value $\tilde V_i$. The coefficients $v_n$ and $\tilde V_i$ are the DOFs defining the scalar potential $v$ in the discrete function space $\vspd(\Om)$.

\subsubsection{Equivalent circuit}

Equations~\eqref{eq_coupling_I} and \eqref{eq_coupling_V} can be either introduced as is, or alternatively implicitly imposed via the definition of an electrical circuit linking the global quantities. The latter option is chosen for the implementation in GetDP. The circuit is defined explicitly as a network of nodes and branches linking distinct filaments in accordance with the permutation operator $\mathcal S$ of the considered geometry. 
The circuit network creates a system of $2\Nf$ equations to be solved together with the equations resulting from the \axial and \transverse formulations.

\section{Verification with Linear Materials}\label{sec_verification}

As a first verification step, we consider a linear problem with the superconducting filaments modelled with very low constant resistivity. We fix the resistivity in the matrix to the constant value $\rho_\text{Cu} = 1.81\times 10^{-10}$ $\Omega$m, and that of the filaments to $\rho_\text{SC} = 10^{-5} \rho_\text{Cu}$. Such a linear model does not allow to describe hysteresis effects in superconducting filaments, but it contains the coupling current physics. We choose this configuration in order to have an efficient reference case (see Section~\ref{sec_helicoidalReference}), to assess the length correction and the coupling current dynamic correction. The case of a nonlinear problem will be discussed in Section~\ref{sec_verification_nonlinear}, as a second verification step.

We consider the 54-filament strand geometry represented in Fig.~\ref{domain_definition}, in which the filaments are arranged in a hexagonal lattice with center-to-center spacing of $110$~$\upmu$m. The filaments diameter is $90$~$\upmu$m, the wire diameter is $d = 1$~mm, and the external air boundary is placed at a distance of $15$ mm from the center of the strand. The twist pitch length of the wire is denoted by $p$. The geometrical parameters of the strand are summarized in Table~\ref{filament54_param}.

\begin{table}[!h]\small
\centering
\begin{tabular}{l r}
\hline
Number of filaments ($\Nf$) & 54\ \ \,\quad ~\\
Filament diameter & $90~\upmu$m\\
Filament center-to-center spacing & $110~\upmu$m\\
Strand diameter ($d$) & $1$~mm\\
Twist pitch length ($p$) & $5$-$100$~mm\\ 
\hline
\end{tabular}
\caption{Geometrical parameters of the 54-filament strand.}
\label{filament54_param}
\end{table}

The transport current $I_\text{t}$ is fixed to zero and the wire is subject to a sinusoidally-varying spatially-uniform transverse magnetic flux density along $\ey$ of amplitude $b = 1$~T and frequency $f$. Parameters $f$ and $p$ are the main parameters of the verification study.

Since the problem is linear, it can be solved in the frequency domain. For this, the \axial model is written in terms of the auxiliary complex quantity $\vec{\hat h}(\vec x)$, the phasor of the magnetic field, with $\vec x$ the position vector. The phasor is related to the physical magnetic field by $\h(\vec x, t) = \Re\big(\vec{\hat h}(\vec x)e^{i\omega t}\big)$, with $i = \sqrt{-1}$ and $\omega = 2\pi f$, and we replace all time derivatives in the formulation by a multiplication by $i\omega$. Similarly, the \transverse model is written in terms of the phasor of the electric scalar potential, $\hat v(\vec x)$, with $v(\vec x, t) = \Re\big(\hat v(\vec x)e^{i\omega t}\big)$.

From the complex solution of the \method method, the time-average power loss (or total AC loss) per unit length $P_\text{tot}$, in W/m, is defined as
\begin{align}\label{eq_ptot_linear}
P_\text{tot} =& \underbrace{\frac{1}{2} \int_{\Of} \rho\, \vec{\hat j}^\star \cdot \vec{\hat j}\,d\O}_{P_\text{filament}}+ \underbrace{\frac{1}{2} \int_{\Om} \rho\, \vec{\hat j}^\star \cdot \vec{\hat j}\,d\O}_{P_\text{eddy}}\notag\\
&+ \underbrace{\frac{1}{2}\int_{\Om} \rho\, \vec{\hat j}_{\text{cc,d}}^\star \cdot \vec{\hat j}_{\text{cc,d}}\,d\O}_{P_\text{coupling}},
\end{align}
with $\vec{\hat j}$ the axial current density phasor and $\vec{\hat j}_{\text{cc,d}}$ the transverse current density phasor (with dynamic correction), and with $\vec{\hat j}^\star$ and $\vec{\hat j}_{\text{cc,d}}^\star$ their complex conjugates, respectively.

The total AC loss $P_\text{tot}$ can be decomposed into the three distinct contributions $P_\text{filament}$, $P_\text{eddy}$, and $P_\text{coupling}$, as defined in Eq.~\eqref{eq_ptot_linear}. Multiplying these quantities by the period $1/f$ gives the AC loss per cycle and per unit length, $Q_\text{tot}$, $Q_\text{filament}$, $Q_\text{eddy}$, and $Q_\text{coupling}$, in J/m.

\subsection{Reference model}\label{sec_helicoidalReference}

As a reference model, we consider the helicoidal transformation method proposed in~\cite{dular2023helicoidal}, using the change of variables introduced in~\cite{nicolet2004modelling} and applied on the \hpfOnly. The method developed in~\cite{dular2023helicoidal} allows for an exact treatment of transverse magnetic fields in the case of linear materials. We refer to this reference method as the helicoidal method, and we also solve it in the frequency domain. The total loss is also decomposed into three components, similarly to Eq.~\eqref{eq_ptot_linear}, as defined in~\cite{dular2023helicoidal}.

We consider the same 2D geometry and mesh of the cross section for the \method and helicoidal methods. The chosen mesh leads to a number of 63k DOFs with the \method method and 113k DOFs with the helicoidal method.

Note that the two models do not define exactly the same physical problem. Indeed, the helicoidal model describes tilted filaments rigorously, with three-dimensional current flow, whereas this is neglected with the \method method, as the filaments are modelled as straight conductors. As a result, the filaments in the helicoidal method have a reduced effective cross section compared to the circular section in the $x$-$y$-plane. For real strands however the importance of this effect is limited. The effective cross section reduction of a filament is of the order of $1-\cos \theta$ with $\theta$ the tilt angle of that filament with respect to the $\ez$ direction. For the typical value $p/d = 20$, $\theta < 9\degree$ and a maximum difference of the order of $1.2$\% is therefore expected locally on the geometry.

\subsection{Results and error map}

In this section, we first consider the \method method including both the corrected length $\ell^{\text{c}}$ and the dynamic correction described in Sections~\ref{sec_correctedLength} and \ref{sec_dynamicCorrection}, respectively. In the next two sections, we then discuss the benefits brought by these two corrections individually.

\begin{figure}[ht!]
    \centering
	\includegraphics[width=\linewidth]{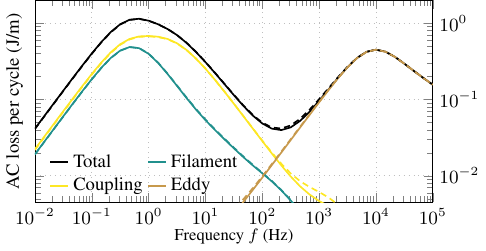}
    \caption{Total AC loss per cycle and separate contributions versus the frequency $f$ of an applied transverse magnetic field (of amplitude $1$~T), for $p/d=20$ and the 54-filament strand with linear material properties. Solid curves are results of the \method method. Dashed curves are results of the reference helicoidal method.}
    \label{linearACloss_pd20}
\end{figure}

The AC loss per cycle and their different contributions for $p/d=20$ are given in Fig.~\ref{linearACloss_pd20}, for both the \method method and the reference helicoidal method. The agreement between the models is excellent at all frequencies.

The \method method reproduces the characteristic bell shape curve of the coupling loss, associated with transverse currents, or coupling currents, flowing in the matrix between the filaments. Analytical models describe the coupling loss per cycle as follows~\cite{campbell1982general}:
\begin{align}\label{eq_timeConstantCoupling}
Q_\text{coupling} = \frac{\pi d^2}{4} \frac{b^2}{2\mu_0}\,\frac{\pi \omega \tau_\text{c}}{(\omega^2 \tau_\text{c}^2 + 1)},
\end{align}
with $\mu_0=4\pi\times 10^{-7}$~H/m and the interfilament coupling time constant $\tau_\text{c}$ defined by
\begin{align}\label{eq_timeConstant}
\tau_\text{c} = \frac{\mu_0}{2\rho_{\text{eff}}} \paren{\frac{p}{2\pi}}^2,
\end{align}
where $\rho_{\text{eff}}$ is an effective resistivity of the matrix, that accounts for the presence of the filaments~\cite{wilson1983superconducting}. Assuming no insulation between the filaments and the matrix~\cite{wilson2008nbti}:
\begin{align}
\rho_\text{eff} = \rho_\text{Cu}\, \frac{1-\lambda}{1+\lambda},
\end{align}
with $\lambda$ the filling factor of the filaments in the wire. With the strand considered here, $\tau_\text{c} = 90$~ms for $p=20$~mm. The associated frequency is $f_\text{c} = (2\pi \tau_\text{c})^{-1} = 1.8$~Hz, which roughly corresponds to the position of the peak of the coupling loss per cycle in Fig.~\ref{linearACloss_pd20}.

The second peak in total AC loss per cycle is related to axial eddy currents, and the associated skin effect. They first increase as $\sim f$, and then decrease as $\sim 1/\sqrt{f}$, with a change of regime arising when the diffusion skin depth $\delta_\text{Cu} = \sqrt{2\rho_\text{Cu} /\omega \mu_0}$ is comparable with the thickness of the outer sheath of the matrix $d_\text{os}~\approx~ 80$~$\upmu$m~\cite{dular2023helicoidal}. Here, we have $\delta_\text{Cu} / d_\text{os} = 1$ for frequency $f = 7.2$~kHz, which is close to the position of the peak of the eddy loss per cycle in Fig.~\ref{linearACloss_pd20}.

\begin{figure}[h!]
\begin{center}
\includegraphics[width=\linewidth]{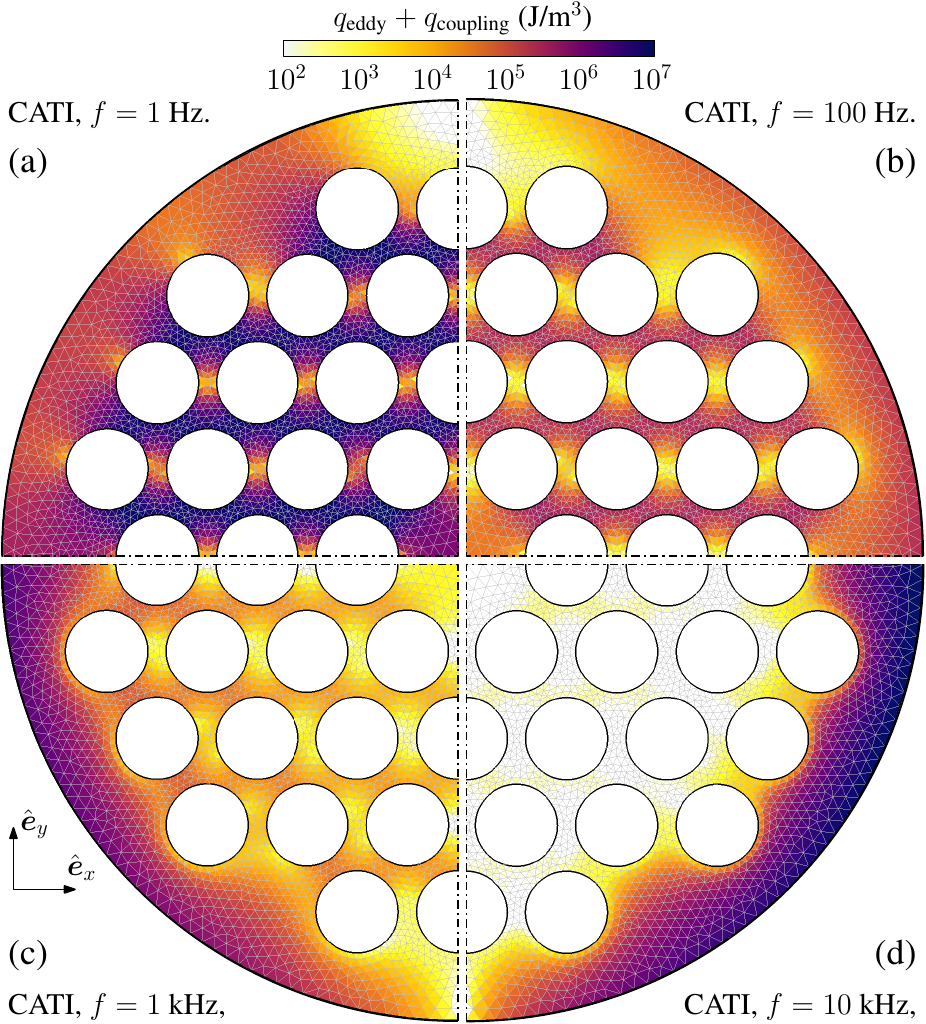}
\caption{Energy loss density per cycle $q_\text{eddy}+q_\text{coupling}$ in the matrix for $p/d=20$ with linear material properties, at four different frequencies. Each case is represented on one quarter of the cross section, dash-dotted lines represent symmetry lines. The colormap is in a log-scale.}
\label{matrix_loss_all_four}
\end{center}
\end{figure}

To further illustrate these regimes and the ability of the \method method to reproduce them, the power loss density per cycle in the matrix, $q_\text{eddy}+q_\text{coupling}$, in J/m$^3$, with
\begin{align}
q_\text{eddy} = \frac{\rho}{2f}\ \vec{\hat j}^\star \cdot \vec{\hat j}, \quad q_\text{coupling} = \frac{\rho}{2f}\ \vec{\hat j}_{\text{cc,d}}^\star \cdot \vec{\hat j}_{\text{cc,d}},
\end{align}
is represented in Fig.~\ref{matrix_loss_all_four} at four different frequencies. It depicts how the energy loss density is progressively pushed towards outer parts of the matrix as the frequency increases.

The shape of the curve describing filament losses is that of linear resistive filaments with their associated skin effect~\cite{dular2023helicoidal}. It therefore does not describe the hysteresis power loss of superconducting filaments, described by a nonlinear resistivity. The response of superconducting filaments is qualitatively different (see, e.g., Fig.~\ref{lossMap_cut_0p2T_0A}), and is discussed in Section~\ref{sec_lossmap}.

\begin{figure}[h!]
        \centering
\centering
	\includegraphics[width=\linewidth]{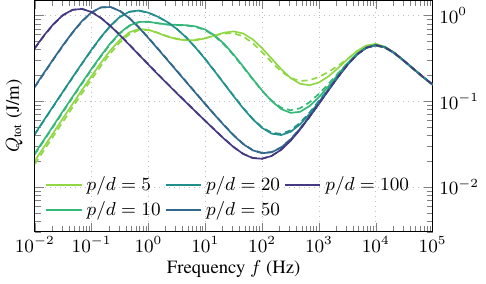}
\caption{Total AC loss per cycle and per unit length as a function of the frequency $f$ of an applied transverse magnetic field, for different twist pitch lengths to diameter ratios $p/d$ for the 54-filament strand with linear material properties. Solid curves are results of the \method method. Dashed curves are results of the reference helicoidal method.}
        \label{linearACloss_p}
\end{figure}

The total AC loss is now compared for different $p/d$ ratios, in Fig.~\ref{linearACloss_p}. The results of the \method method match very well with the reference results in most cases. The difference between both methods becomes noticeable for low $p/d$ ratios only ($p/d \lesssim 10$), which is expected. Indeed, for $p/d=5$, the tilt angle of filaments in the outer layer is $26\degree$, so that $1-\cos(26\degree) \approx 10\%$, and the validity of describing the geometry with 2D models is questionable.

\begin{figure}[h!]
\begin{center}
\includegraphics[width=\linewidth]{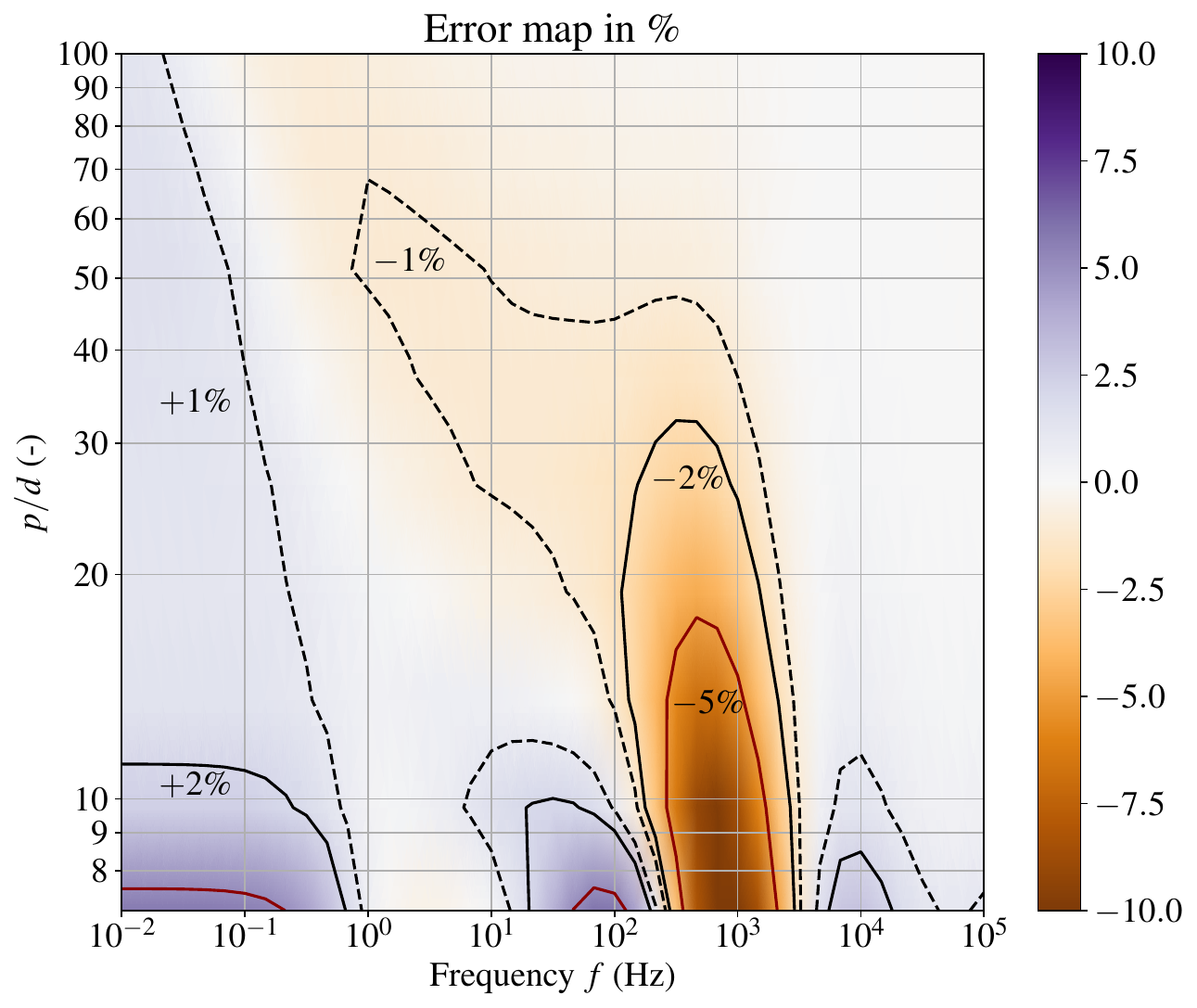}
\caption{Relative difference between the total AC loss predicted by the \method method and the reference values of the helicoidal method, as a function of the frequency $f$ of an applied transverse magnetic field, and of the twist pitch length to diameter ratio $p/d$ of the 54-filament strand with linear material properties. Dashed curves are $\pm 1\%$ contour lines, solid black curves are $\pm 2\%$ contour lines and solid red curves are $\pm 5\%$ contour lines.}
\label{linearErrorMap_dyn_ellstar}
\end{center}
\end{figure}

In order to quantify the difference between the \method and helicoidal methods, the relative difference between their results is represented in Fig.~\ref{linearErrorMap_dyn_ellstar}, as a function of the frequency $f$ and the twist pitch length over diameter ratio $p/d$. The largest error is observed for small $p/d$ ratios, in the $0.2-2$ kHz range. For $p/d\approx 20$ and above, the relative error is smaller than $5\%$ in the entire frequency range.

\subsection{Discussion of the length correction factor}

\begin{figure}[h!]
        \centering
\centering
	\includegraphics[width=\linewidth]{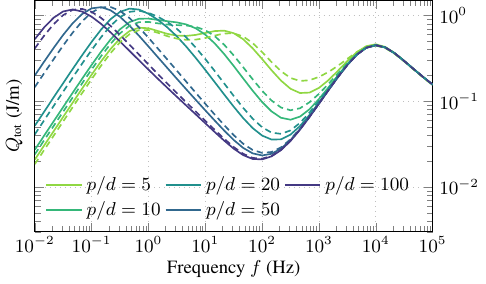}
\caption{Effect of the length correction factor. Solid curves are results of the \method method without the length correction, i.e., using $\ell$ and not $\ell^\text{c}$ in Eqn.~\eqref{eq_oop} and \eqref{eq_ip}. Dashed curves are reference results of the helicoidal method.}
        \label{linearErrorMap_dyn_ell}
\end{figure}

The corrected length $\ell^\text{c}$ was introduced in Section~\ref{sec_correctedLength} in order to correct for an overestimation of the integrated quantities along $\ez$. Using the periodicity length $\ell$ instead of the corrected length $\ell^\text{c}$ leads to much higher error compared to the reference solution of the helicoidal method. This is illustrated in Fig.~\ref{linearErrorMap_dyn_ell}.

The use of $\ell$ instead of $\ell^\text{c}$ is equivalent to overestimating the twist pitch length $p$, and hence the time constant $\tau_\text{c} = (2\pi f_\text{c})^{-1}$ associated with the coupling currents, Eq.~\eqref{eq_timeConstant}. With Eq.~\eqref{eq_timeConstantCoupling}, one can see that this shifts the coupling current loss curve towards lower frequencies, leading to loss overestimation for $f<f_\text{c}$ and underestimation for $f>f_\text{c}$. At higher frequencies (for $f\gtrsim 3$ kHz for this wire), the eddy current loss associated with axial currents in the outer sheath of the matrix starts to dominate, so that the contribution of the coupling loss to the total loss decreases. This reduces the error in Fig.~\ref{linearErrorMap_dyn_ell}, but the error in the coupling loss is still there.

\subsection{Correction for dynamic effects in the matrix}\label{sec_dynCorrEffect}

\begin{figure}[h!]
        \centering
	\includegraphics[width=\linewidth]{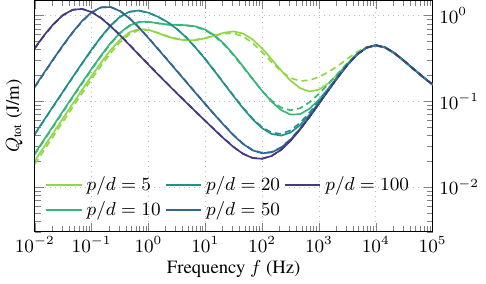}
\caption{Effect of the dynamic correction. Solid curves are results of the \method method without the dynamic correction. Dashed curves are reference results of the helicoidal method.}
        \label{linearErrorMap_nodyn_ellstar}
\end{figure}

The dynamic correction introduced in Section~\ref{sec_dynamicCorrection} is a proposal to account for eddy current effects on the transverse coupling current density distribution. Without this correction, we observe a higher difference between the \method method and the reference helicoidal solution in the $0.1-10$ kHz frequency range, as shown in Fig.~\ref{linearErrorMap_nodyn_ellstar}. 

\begin{figure}[h!]
\begin{center}
\includegraphics[width=\linewidth]{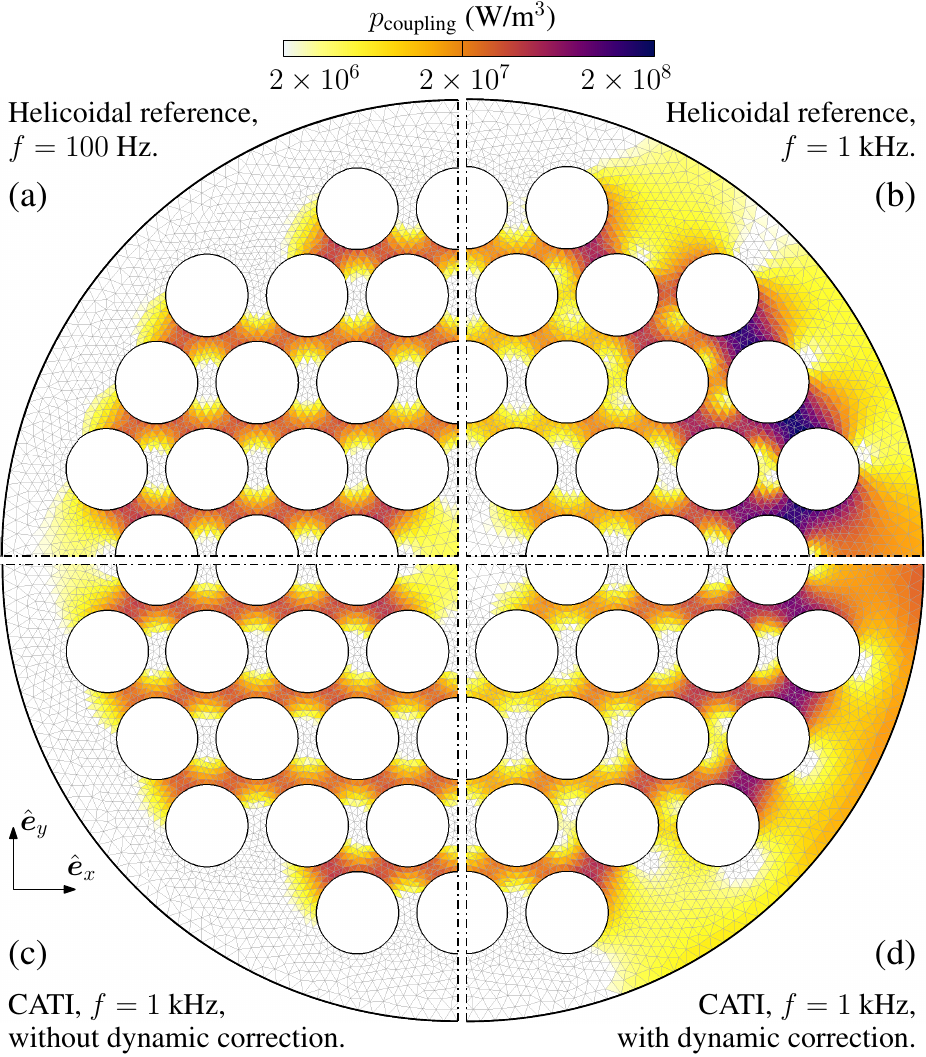}
\caption{Power loss density $p_\text{coupling}$ due to transverse coupling currents in harmonic regime for $p/d=20$ in four different cases with linear material properties. Each case is represented on one quarter of the cross section, dash-dotted lines represent symmetry lines. The colormap is in a log-scale. (a-b) Reference helicoidal solutions on the plane $z = 0$, at $f=100$~Hz and $f=1$~kHz, respectively. (c-d) \method method solutions at $f=1$~kHz, without and with the dynamic correction, respectively.}
\label{couplingCurrent_dynamics_all_four}
\end{center}
\end{figure}

To illustrate the effect of the dynamic correction, the average power loss density generated by the coupling currents in the strand matrix is shown in Fig.~\ref{couplingCurrent_dynamics_all_four}(a-b) for two frequencies, using the reference helicoidal method, for $p/d=20$. The time-average power loss density, in W/m$^3$, is computed as 
\begin{align}
p_\text{coupling} = \frac{1}{2}\,\vec{\hat j}^\star_{\text{cc}}\cdot(\rho\, \vec{\hat j}_{\text{cc}}),
\end{align}
with $\vec{\hat j}_{\text{cc}}$ the transverse current density phasor and $\vec{\hat j}^\star_{\text{cc}}$ its complex conjugate. For the reference helicoidal method, $\vec{\hat j}_\text{cc}$ is the projection of the full current density phasor on the $x$-$y$-plane, in the Cartesian coordinate system. At the frequency $f=100$ Hz, the coupling currents are distributed uniformly in-between the filaments, whereas at $f=1$ kHz, they are mostly localized in the outer part of the strand, with non-negligible contributions in the outer sheath of the matrix.

Without the dynamic correction, the \method method reproduces accurately the coupling current distribution at low frequencies, e.g., at $f=100$ Hz, when the current flow is mostly a curl-free field. However, this is not the case at higher frequencies, e.g., at $f=1$ kHz, as illustrated in Fig.~\ref{couplingCurrent_dynamics_all_four}(c). Correcting the current flow with the contribution from a dynamic magnetic field $\hd$ leads to redistributed coupling currents $\jccdyn$ that are closer to reality, as shown in Fig.~\ref{couplingCurrent_dynamics_all_four}(d). Still, as can be seen from the error map in Fig.~\ref{linearErrorMap_dyn_ellstar}, the match is not perfect. The proposed dynamic correction is only approximate. Improving it with 2D models only is however not obvious because the coupling current dynamics is a 3D effect in essence.

It is important to stress that the dynamic correction has a noticeable effect on the total loss in a limited range of frequencies only. When the frequency is high enough for the axial skin effect to dominate ($f \gtrsim 3$ kHz, here), the axial eddy currents dominate the total loss, and the contribution of the coupling losses to the total loss, with or without dynamic correction, becomes negligible. 

\section{Verification with Nonlinear Materials}\label{sec_verification_nonlinear}

As a second verification step, we consider the 54-filament strand geometry defined in Table~\ref{filament54_param}, with $p~=~19$ mm, but now with realistic field-dependent material properties for both the Nb-Ti filaments and the copper matrix\footnote{A version of GetDP compiled with these material functions can be found online at \url{https://cern.ch/cerngetdp}.}. We fix the temperature to $1.9$~K. The resistivity of the Nb-Ti filaments is described by the power law~\cite{rhyner1993magnetic}:
\begin{equation}\label{eqn_contitutiveje}
\rho_\text{SC}(\j,\b) = \frac{\ec}{\jc(\|\b\|)}\paren{ \frac{\|\j\|}{\jc(\|\b\|)}}^{n-1},
\end{equation}
with $\ec = 10^{-4}$ V/m, $n=30$, and a field-dependent critical current density $\jc(\|\b\|)$ (A/m$^2$) obtained from the STEAM material library~\cite{zachou2024}. The resistivity of the copper matrix, $\rho_\text{Cu}(\b)$, accounting for magneto-resistance, is also taken from~\cite{zachou2024}, with residual resistivity ratio $\text{RRR} = 100$. The magnetic flux density $\b = \mu_0 \h$ in the material parameter functions is expressed from the magnetic field $\h$ of the \axial model. This introduces an additional coupling between both models.

The strand is subject to a uniform transverse magnetic flux density field along $\ey$, varying sinusoidally with an amplitude $b$ and a frequency $f$. The transport current is fixed to zero.

As the problem is now nonlinear, it can no longer be solved in the frequency domain. An implicit (backward) Euler time-stepping is used for time integration~\cite{griffiths2010numerical}. In addition, an iterative algorithm has to be used in order to solve the resulting nonlinear system of equations at each time step. A Newton-Raphson technique is chosen to this end, as it leads to efficient resolutions for problems involving the power law resistivity~\cite{dular2019finite}.

The instantaneous AC loss per unit length, $P_\text{tot}$ (W/m), is evaluated as follows:
\begin{align}\label{eq_q_tot_time}
P_\text{tot} =& \underbrace{\int_{\Of} \rho\, \j \cdot \j\,d\O}_{P_\text{filament}} + \underbrace{\int_{\Om} \rho\, \j \cdot \j\,d\O}_{P_\text{eddy}}+ \underbrace{\int_{\Om} \rho\,  \jcc \cdot \jcc\,d\O}_{P_\text{coupling}}.
\end{align}
It is decomposed in three distinct contributions, defined by the three terms of Eq.~\eqref{eq_q_tot_time}, as was done in Eq.~\eqref{eq_ptot_linear}. This decomposition for AC loss components is typical in the literature and helps to better interpret the simulation results.

The AC loss per cycle per unit length is defined as the loss per cycle of applied field, $Q_\text{tot}$, in J/m. Numerically, it is evaluated as follows~\cite{shen2020review}, with $T = 1/f$,
\begin{align}
Q_\text{tot} = 2\int_{T/2}^{T} P_\text{tot}(t)\, dt.
\end{align}

\subsection{Reference model}

As a reference, we consider a classical 3D FE model with the \hpfOnly~\cite{dular2023helicoidal}, as represented in Fig.~\ref{3Dgeo_54fil_mesh_full}. Periodic boundary conditions are imposed on the bases of the cylindrical wire so that only a length of $p/6$ needs to be modelled. A structured prismatic mesh is used in the filaments, whereas an unstructured mesh is used in the matrix and air domains. To resolve the potentially small penetration depth of the induced currents, a sufficiently fine spatial discretization is necessary, which leads to a high computational cost, compared to the \method method.

\subsection{Comparison and validity of the \method method}

The \method method and the reference 3D method are compared at a field amplitude of $b=0.5$~T at frequencies of $f=10$~Hz and $f=1$~kHz. These conditions represent non-trivial solutions where the filament loss is of the same order of magnitude as the coupling loss for $f~=~10$~Hz, and as the eddy loss for $f~=~1$~kHz. The chosen field amplitude also avoids computational challenges associated with both low and high field amplitudes in the 3D reference model, where low fields require substantial mesh refinement to capture the field penetration close to the surface of the filaments, and high fields face convergence difficulties associated with the non-linear material properties, requiring substantial time step refinement. Both the corrected length $\ell^{\text{c}}$ and the dynamic correction are included in the \method method.

\begin{figure}[ht!]
    \centering    
    \begin{subfigure}[b]{\linewidth}
        \centering
        \includegraphics[width=\linewidth]{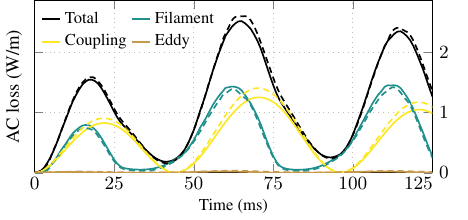}
        \caption{$f=10$ Hz.}
        \label{fig:3D_verif_AC_loss_f10}
    \end{subfigure}
    \vspace{0.01cm} 
    \begin{subfigure}[b]{\linewidth}
        \centering
        \includegraphics[width=\linewidth]{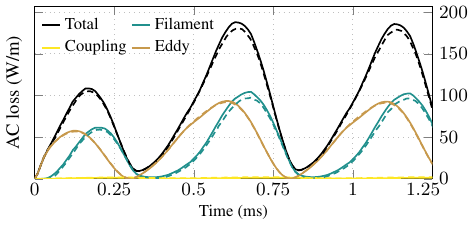}
        \caption{$f=1$ kHz.}
        \label{fig:3D_verif_AC_loss_f1000}
    \end{subfigure}
    \caption{Comparison of the AC losses in the composite strand obtained from the \method method (solid lines) and the reference 3D model (dashed lines). The strand is subject to $1.25$ cycles of a sinusoidally varying transverse magnetic flux density, of amplitude $0.5$ T at frequencies of (a) $10$~Hz and (b) $1$~kHz.}
    \label{fig:3D_verif_AC_loss}
\end{figure}

\begin{figure}[h!]
\begin{center}
\includegraphics[width=\linewidth]{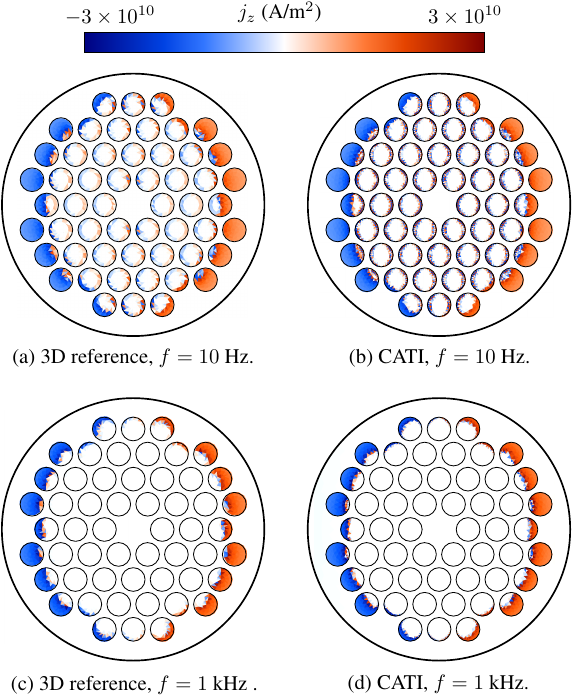}
\caption{Comparison of the axial current density distributions at the last time step of the simulations, $t=1.25/f$, for the \method method and the reference 3D model (at $z=0$), for a field amplitude of 
$0.5$~T and a frequency of (a-b) $f=10$~Hz and (c-d) $f=1$~kHz.}
\label{fig:3D_verif_solutions}
\end{center}
\end{figure}

Figure \ref{fig:3D_verif_AC_loss} shows comparisons of the computed AC losses between the \method method (solid lines) and the reference 3D method (dashed lines) over the span of $1.25$ cycles of the applied field, with frequencies of $10$~Hz and $1$~kHz. At the $10$~Hz frequency, the relative error on the total AC losses per cycle is $-3.44\%$, with contributions from the filament, coupling, and eddy losses of $+3.05\%$, $-5.9\%$, and $-0.59\%$, respectively. At the $1$~kHz frequency, the relative error on the total loss is $+2.38\%$, with contributions from the filament, coupling, and eddy losses of $+3.17\%$, $-0.82\%$, and $+0.03\%$, respectively. Figure \ref{fig:3D_verif_solutions} provides a qualitative comparison of the axial current density distributions in the wire at time $t=1.25/f$, for the \method method and the reference method, at $10$~Hz and $1$~kHz. Both locally and globally, the \method method produces accurate results.

A significant difference in computational time is observed between the 3D reference model, with 970k DOFs, and the \method method, with 28.5k DOFs. These simulations each took more than a week ($168$ h) to complete for the reference 3D model, while they only took about $1$ h with the \method method. The reference model is also quickly limited in terms of mesh refinement and by convergence difficulties. This is not the case with the \method method.

Considering a transport current in addition to the transverse field does not bring new difficulties. The \method method produces reliable results in that case as well.

\section{Applications}\label{sec_applications}

In this section, we demonstrate two applications of the \method method. The first application consists in the analysis of the response of a helicoidally symmetric strand to transverse fields of various frequencies and amplitudes, in terms of power loss and magnetization. The low computational cost of the \method method compared to that of full 3D models, enables to perform detailed parameter sweep studies in a reasonable computational time, and with a reasonable amount of computational resources.

The second application consists in the computation of losses in a wire-in-channel geometry subject to DC transport current and magnetic field, under the application of external AC field ripples. This application illustrates the possibility of modelling periodic, but non-helicoidally symmetric conductor geometries with the \method method.

We consider nonlinear materials for both applications.

\subsection{Loss map and magnetization curves}\label{sec_lossmap}

\begin{figure}[h!]
\begin{center}
\includegraphics[width=\linewidth]{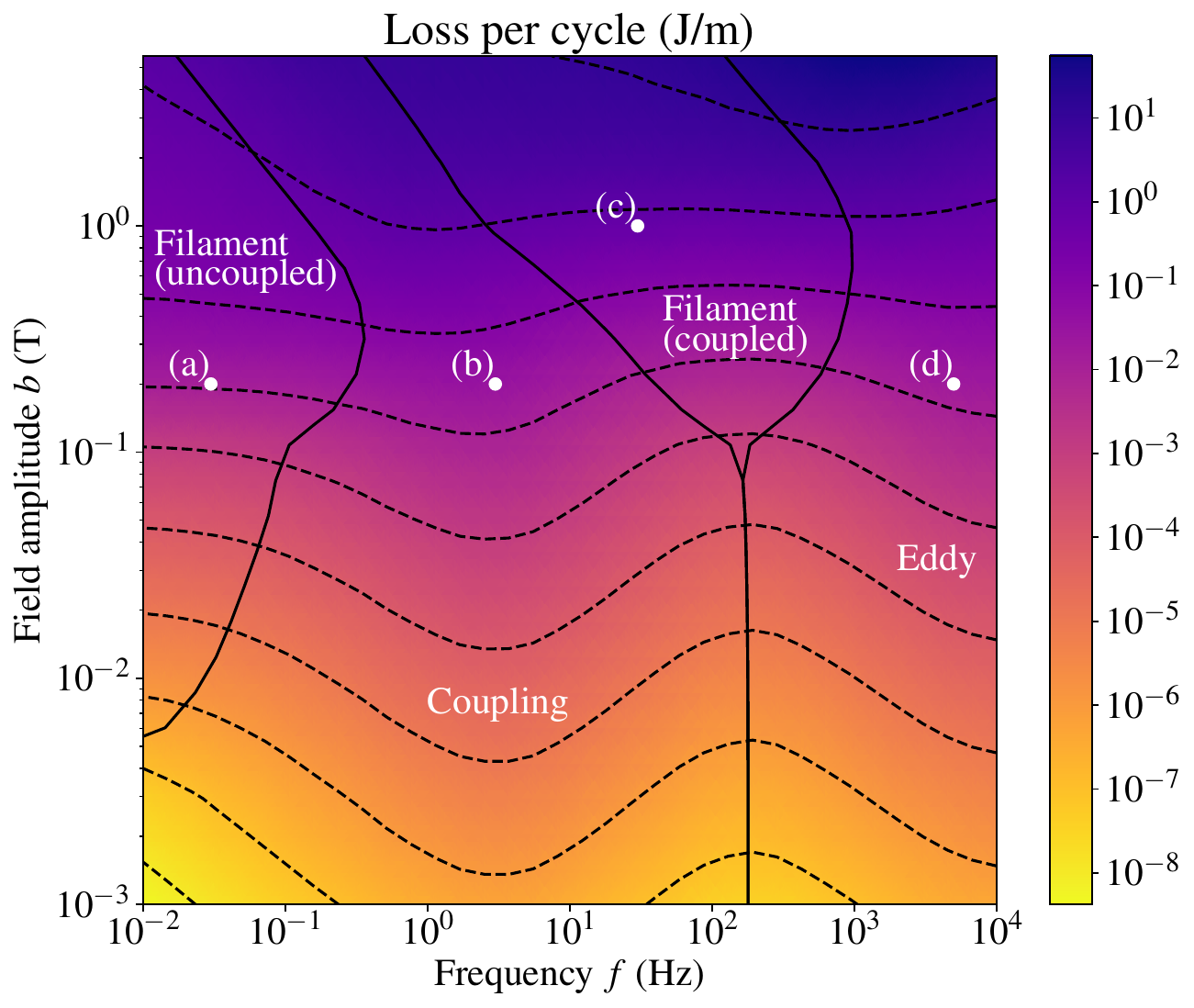}
\caption{Total loss per cycle and per unit length as a function of the frequency and amplitude of an applied transverse field, and no transport current. Dashed curves are contour lines of equal loss. Solid curves delimit areas of distinct dominant loss contributions (filament, coupling, or eddy losses). The four points are at positions where the solution is represented in Fig.~\ref{cross_sections_j}(a-d). This loss map is inspired from the work of A. M. Campbell in~\cite{campbell1982general}.}
\label{loss_map_I_0}
\end{center}
\end{figure}

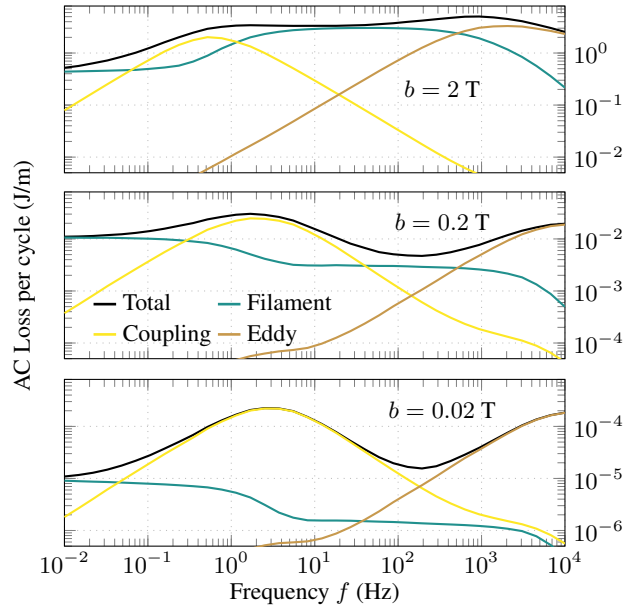
\begin{figure}[h!]
\centering
 \begin{subfigure}[b]{0.99\linewidth}  
 \centering
\begin{tikzpicture}[trim axis left, trim axis right][font=\small]
\pgfplotsset{set layers}
 	\begin{loglogaxis}[
	tick label style={/pgf/number format/fixed},
    width=\linewidth,
    height=3.8cm,
    grid = major,
    grid style = dotted,
    ymin=5e-3, 
    ymax=8,
    xmin=0.01, 
    xmax=10000,
    xticklabels={},
    ylabel={\phantom{AC pcy(J/m)}},
    ylabel style={yshift=-2.8em},
    xlabel style={yshift=0.5em},
    xticklabel style={yshift=0.1em},
    yticklabel style={xshift=0em},
    yticklabel pos=right,
    legend columns=2,
    legend style={at={(0.38, 0.02)}, cells={anchor=west}, anchor=south, draw=none,fill opacity=0, text opacity = 1}
    ]
    \addplot[black, thick] 
    table[x=f,y=total]{data/strand_lossCycle_2p0T.txt};
        \addplot[vir_3, thick] 
    table[x=f,y=fil]{data/strand_lossCycle_2p0T.txt};
        \addplot[vir_0, thick] 
    table[x=f,y=coupling]{data/strand_lossCycle_2p0T.txt};
        \addplot[myorange, thick] 
    table[x=f,y=eddy]{data/strand_lossCycle_2p0T.txt};
\node[anchor=south] at (axis cs: 337, 0.1) {$b = 2$ T};
    \end{loglogaxis}
\end{tikzpicture}%
\end{subfigure}
        \hfill\vspace{-0.5cm}
 \begin{subfigure}[b]{0.99\linewidth}  
 \centering
\begin{tikzpicture}[trim axis left, trim axis right][font=\small]
\pgfplotsset{set layers}
 	\begin{loglogaxis}[
	tick label style={/pgf/number format/fixed},
    width=\linewidth,
    height=3.8cm,
    grid = major,
    grid style = dotted,
    ymin=5e-5, 
    ymax=8e-2,
    xmin=0.01, 
    xmax=10000,
    xticklabels={},
    ylabel={AC Loss per cycle (J/m)},
    ylabel style={yshift=-2.2em},
    xlabel style={yshift=0.5em},
    xticklabel style={yshift=0.1em},
    yticklabel style={xshift=0em},
    yticklabel pos=right,
    legend columns=2,
    legend style={at={(0.3, 0.02)}, cells={anchor=west}, anchor=south, draw=none,fill opacity=0, text opacity = 1, legend image code/.code={\draw[##1,line width=1pt] plot coordinates {(0cm,0cm) (0.3cm,0cm)};}}
    ]
    \addplot[black, thick] 
    table[x=f,y=total]{data/strand_lossCycle_0p2T.txt};
        \addplot[vir_3, thick] 
    table[x=f,y=fil]{data/strand_lossCycle_0p2T.txt};
        \addplot[vir_0, thick] 
    table[x=f,y=coupling]{data/strand_lossCycle_0p2T.txt};
        \addplot[myorange, thick] 
    table[x=f,y=eddy]{data/strand_lossCycle_0p2T.txt};
    \legend{Total, Filament, Coupling, Eddy}
    \node[anchor=south] at (axis cs: 337, 0.01) {$b = 0.2$ T};
    \end{loglogaxis}
\end{tikzpicture}%
\end{subfigure}
        \hfill\vspace{-0.3cm}
 \begin{subfigure}[b]{0.99\linewidth}  
        \centering
\begin{tikzpicture}[trim axis left, trim axis right][font=\small]
\pgfplotsset{set layers}
 	\begin{loglogaxis}[
	tick label style={/pgf/number format/fixed},
    width=\linewidth,
    height=3.8cm,
    grid = major,
    grid style = dotted,
    ymin=5e-7, 
    ymax=8e-4,
    xmin=0.01, 
    xmax=10000,
	xlabel={Frequency $f$ (Hz)},
    ylabel={\phantom{AC pcy(J/m)}},
    ylabel style={yshift=-2.8em},
    xlabel style={yshift=0.5em},
    xticklabel style={yshift=0.1em},
    yticklabel style={xshift=0em},
    yticklabel pos=right,
    legend columns=2,
    legend style={at={(0.38, 0.03)}, cells={anchor=west}, anchor=south, draw=none,fill opacity=0, text opacity = 1, legend image code/.code={\draw[##1,line width=1pt] plot coordinates {(0cm,0cm) (0.3cm,0cm)};}}
    ]
    \addplot[black, thick] 
    table[x=f,y=total]{data/strand_lossCycle_0p02T.txt};
        \addplot[vir_3, thick] 
    table[x=f,y=fil]{data/strand_lossCycle_0p02T.txt};
        \addplot[vir_0, thick] 
    table[x=f,y=coupling]{data/strand_lossCycle_0p02T.txt};
        \addplot[myorange, thick] 
    table[x=f,y=eddy]{data/strand_lossCycle_0p02T.txt};
\node[anchor=south] at (axis cs: 337, 0.0001) {$b = 0.02$ T};
    \end{loglogaxis}
\end{tikzpicture}%
\end{subfigure}
 \hfill
\vspace{-0.2cm}
\caption{Total loss per cycle and separated loss contributions as a function of the frequency $f$, for three different field amplitudes. The legend is the same for the three subfigures.}
        \label{lossMap_cut_0p2T_0A}
\end{figure}

As a first application of the \method method, we consider the 54-filament strand (see Table~\ref{filament54_param}) with the same material properties as in Section~\ref{sec_verification_nonlinear}. We model its response to an external transverse magnetic field, varying sinusoidally with frequencies $f$ ranging from $0.01$ Hz to $10$ kHz, sampled with $25$ values, and amplitudes $b$ from $0.001$ T to $5.6$ T, sampled with $16$ values. We fix the twist pitch length to $p=19$ mm. A loss map, inspired by the work of A. M. Campbell in~\cite{campbell1982general}, is plotted in Fig.~\ref{loss_map_I_0} and gives the total AC loss per cycle and per unit length of the strand, interpolated from the results ($400$ simulations) in this parameter space. The AC loss per cycle is computed as described in Section~\ref{sec_verification_nonlinear}. The different loss contributions are also illustrated as a function of frequency in Fig.~\ref{lossMap_cut_0p2T_0A} for field amplitudes of $0.02$~T, $0.2$~T, and $2$~T.

\begin{figure}[h!]
\begin{center}
\includegraphics[width=\linewidth]{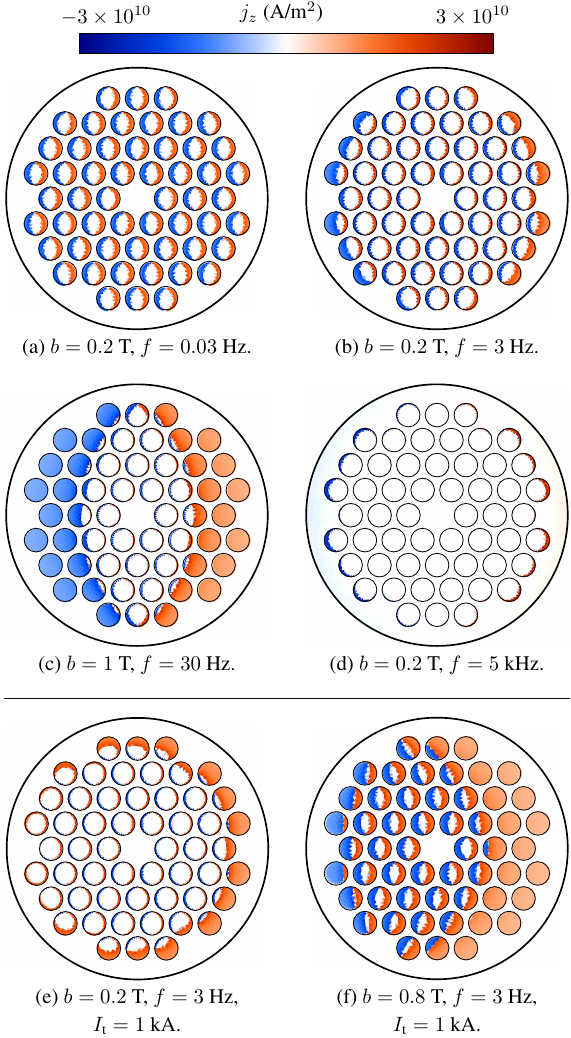}
\caption{Axial current density distributions in $\Oc$, at time $t=1/(4f)$ for various excitations. The colormap is the same for all subfigures. (a-d) Vertical transverse field with zero transport current, see the four points in Fig.~\ref{loss_map_I_0}. (e-f) Vertical transverse field with transport current in phase with the field.}
\label{cross_sections_j}
\end{center}
\end{figure}

The black lines in Fig.~\ref{loss_map_I_0} delimit four distinct regions in the map, with distinct dominant loss contributions. Fig.~\ref{cross_sections_j}(a-d) gives the current density distribution at specific locations in these four regions.

The frequency dependence of the coupling and eddy loss contributions in Fig.~\ref{lossMap_cut_0p2T_0A} is comparable to that obtained with linear materials in Section~\ref{sec_verification}. The coupling loss follows a typical bell curve~\cite{campbell1982general}, and the shape of the eddy loss curve is associated with skin effect in the matrix. A good match with the analytical model is obtained at low field amplitudes ($b\lesssim 0.1$~T). At higher field amplitudes, saturation effects in the filaments affect the dynamics, which is no longer well described by the analytical model.

The filament loss curve of superconducting (nonlinear) filaments is qualitatively different from that of resistive (linear) filaments. In the nonlinear case, the evolution of the filament loss with frequency can be decomposed in four regimes, see Fig.~\ref{cross_sections_j}(a-d). First, at the lowest frequencies, filaments are fully uncoupled and exhibit hysteresis losses that are, per cycle, almost frequency-independent. As the frequency further increases, coupling currents start to progressively couple the filaments, which leads to a smooth transition of the filament loss per cycle towards a new plateau, during which the coupling loss dominates the total loss. When filaments are fully coupled, the filament loss per cycle is again almost frequency-independent. Whether the filament loss for coupled filaments is larger or smaller than the filament loss for uncoupled filaments depends on the external field amplitude. Finally, at the highest frequencies, eddy currents in the matrix start to shield the inner part of the strand, leading to a sharp decrease of filament losses, because the field seen by the superconducting filaments is decreasing. In this regime, eddy current loss dominates the total loss.

\begin{figure}[h!]
\includegraphics[width=\linewidth]{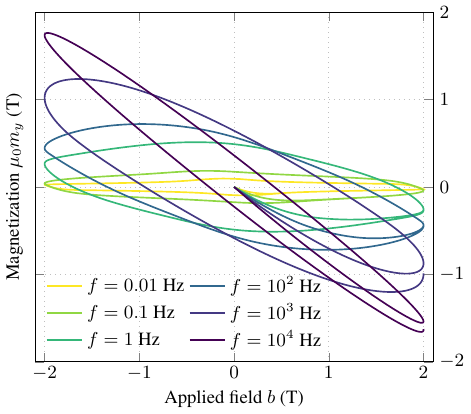}
\caption{Total strand magnetization for a sinusoidally varying external magnetic field of amplitude $2$~T, over $1.25$ periods of the field for different frequencies $f$.}
    \label{magnCurves_2T}
\end{figure}

The strand magnetization is another interesting output of the \method method. In the case with no transport current, the average magnetization vector $\vec m$ (A/m) is defined as~\cite{bossavit2000remarks}
\begin{align}
\vec m = \frac{1}{2S}\int_{\Oc} \vec x \times \j\, \text{d}\Oc,
\end{align}
with $S = \pi d^2/4$ the surface area of the strand and $\vec x$ the position vector. The $y$-component $m_y$ of the magnetization vector $\vec m$ is represented in Fig.~\ref{magnCurves_2T} for different frequencies and an applied transverse field amplitude of $2$ T.

As the frequency increases, the evolution of the curves illustrates the evolution from uncoupled to coupled filaments, with increasing magnetization (for $b=2$~T), and then to almost fully shielded strand due to eddy currents in the matrix, in which case the magnetization loop approaches the shape of an ellipse~\cite{bossavit2000remarks}.

For illustration, solutions with a superimposed transport current $I_\text{t}(t) = 1~\sin(2\pi f t)$ kA, in phase with the applied tranverse field, are shown in Fig.~\ref{cross_sections_j}(e-f). The resulting current density distribution is asymmetric. Some filaments now carry part of the transport current at current density close to the critical current density. This affects the distribution of the magnetization currents in the filaments.

The ability of the \method method to model many different configurations in a reduced amount of time is useful in the context of homogenization methods. The idea for such methods is to replace detailed geometries, e.g., a cable made of many multifilamentary strands, or a magnet made of many such cables, by simplified geometries with appropriate homogenized material parameters reproducing the macroscopic response of the detailed models without describing the details of them explicitly~\cite{niyomsatian2020closed, klop2024electro}. The identification of these material parameters requires the knowledge of the detailed models response under a range of different excitations. The \method method provides an efficient way of computing this accurately.

\subsection{Wire-in-channel geometry}

\begin{figure}[h!]
\begin{center}
\includegraphics[width=\linewidth]{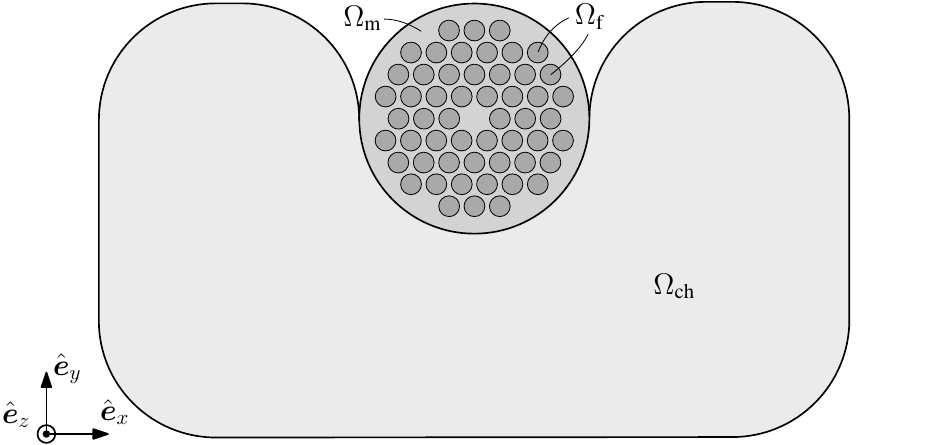}
\caption{Wire-in-channel geometry: copper channel $\Och\subset \Oc$, copper matrix $\Om$, and Nb-Ti filaments $\Of$. The wire $\Of\cup\Om$ is twisted but the channel is translationally invariant along $\ez$.}
\label{wire_in_channel_geo}
\end{center}
\end{figure}

The second application is a conductor with a periodic, but non-helicoidally symmetric, geometry. It consists of the wire-in-channel (WIC) geometry represented in Fig.~\ref{wire_in_channel_geo}. The composite superconducting strand, consisting of matrix and filaments ($\Om~\cup~\Of$), is composed of 54 Nb\nobreakdash-Ti filaments, arranged in a hexagonal lattice with center-to-center spacing of $88$~$\upmu$m. Filament diameter is $72$~$\upmu$m, copper matrix diameter is $d~=~0.8$~mm. The strand is twisted, with a twist pitch length of $p~=~50$~mm, and is placed in a straight copper channel ($\Och$), whose height and width are $1.5$ mm and $2.6$ mm, respectively, with rounded corners of radius $0.4$ mm. The circular numerical air boundary is placed at a distance of $8$~mm from the center of the strand. 

Materials are described by the same resistivities as in the previous section, but with $\text{RRR}=100$ in the channel and $\text{RRR}=200$ in the wire matrix. We assume a high contact resistance between the strand and the channel due to resistive solder in between, which prevent coupling currents to flow from the strand to the channel. The WIC is subject to a uniform background magnetic field of $4$~T along $\ey$ and a DC transport current $I_\text{t} = 500$~A. In addition to the constant background field, a sinusoidal magnetic field of amplitude $b$ and frequency $f$ is applied, either along $\ex$, or along $\ey$.

The simulation is conducted in two steps. In the first step, an initial steady-state solution is computed with a ramp-up during $1$~s of the transport current and background magnetic field, from virgin initial state to $500$~A and $4$~T, respectively. The situation is then held constant for a duration of $4$~s for all dynamic effects to fade out, and the final magnetic field distribution is saved. In the second step, transient simulations are performed with the AC transverse field, using the saved magnetic field distribution from the first step as the initial condition. The AC field has an amplitude of $b=100$~mT and covers the frequency range from $f=0.01$~Hz to $f=10$~kHz. 

\begin{figure}[ht!]
    \centering
    \includegraphics[width=\linewidth]{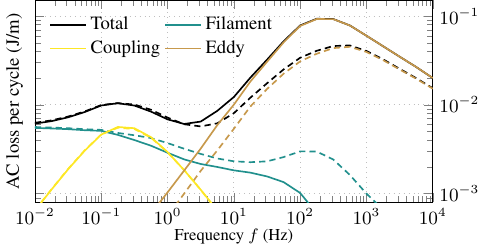}
    \caption{AC loss per cycle in the wire-in-channel geometry as a function of the frequency of the AC field of amplitude $b=100$~mT along $\ex$ (dashed lines) and along $\ey$ (solid lines).}
    \label{fig:wire_in_channel_loss_vs_freq_b0.1}
\end{figure}

\begin{figure}[h!]
\begin{center}
\includegraphics[width=0.86\linewidth]{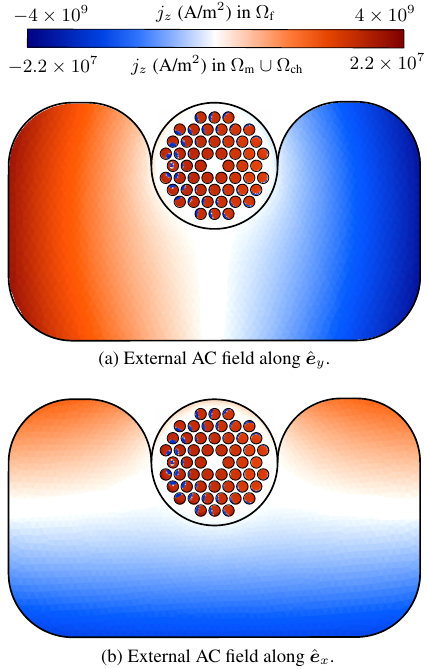}
\caption{Axial current density distributions in the wire-in-channel geometry at $t=1/(2f)$ for the AC fields (a) along $\ey$ and (b) along $\ex$, at $b=100$~mT and $f=100$ Hz. The scale of the colorbars are different in $\Of$ and in $\Om\cup \Och$.}
\label{fig:wire_in_channel_solutions}
\end{center}
\end{figure}

Figure \ref{fig:wire_in_channel_loss_vs_freq_b0.1} shows the loss per cycle due to the AC transverse field as a function of its frequency. Loss contributions are calculated as described in Section~\ref{sec_verification_nonlinear}, with the eddy current loss being integrated over $\Om\cup\Och$. At high frequencies, the eddy current loss dominates and the eddy currents have a screening effect on the filaments, reducing the filament losses. This effect is more pronounced when the AC field is applied along $\ey$, illustrating the anisotropic AC loss response of this non-helicoidally symmetric geometry of WIC. Figure \ref{fig:wire_in_channel_solutions} shows the axial current density distributions in the wire for the two field orientations for $f=100$ Hz at $t=1/(2f)$.

\section{Conclusion}

In this contribution, we proposed the Coupled Axial and Transverse currents (I) (CATI) method. It is based on a pair of coupled 2D models for modelling transient electric and magnetic effects in composite superconducting conductors with a periodic geometry. We applied the \method method to twisted superconducting strands and verified its accuracy by comparison against reference methods. We demonstrated that the \method method reproduces the coupling current density and all types of losses with a very good accuracy in a wide range of frequencies, in the case where the twist pitch length is large enough compared to the strand diameter ($p/d\gtrsim 10$).

As the \method method is based on 2D models, it leads to a massive reduction of the computational cost compared to conventional 3D models, from more than a week (168h) to less than an hour for the tested cases. In particular, this allows to perform detailed parameter sweep studies and in-depth analyses of the strand response in many different configurations in a reasonable amount of time, and with a reasonable amount of computational resources. As an application, such parameter studies could be exploited in the context of homogenization methods, if they are used as detailed mesoscale models for identifying equivalent material parameters of homogenized macroscale models~\cite{niyomsatian2020closed}.

We presented the practical implementation of the \method method, and highlighted the ease of it. Any FE software allowing for field-circuit coupling is in principle ready for its implementation. The use of circuit equations for coupling the \axial and \transverse models also allows for a direct extension of the method for the treatment of contact resistance between the filament and the matrix, as well as that of diffusion barriers in Nb$_3$Sn wires~\cite{hopkins2023design}, that may be modelled as thin-shells~\cite{schnaubelt2023electromagnetic}.

Compared to the helicoidal transformation technique, in addition of handling much more easily nonlinear materials, the \method method also applies to non-helicoidally symmetric cross sections, such as deformed geometries~\cite{hopkins2023design}. Furthermore, an extension of the approach to other types of conductors with periodic geometry, e.g., cables with transposed conductors such as Roebel cables~\cite{goldacker2014roebel}, or Rutherford cables~\cite{verweij1997electrodynamics}, will be subject of further research.

\section*{Acknowledgment}
The work of E. Schnaubelt has been sponsored by the Wolfgang Gentner Programme of the German Federal Ministry of Education and Research (grant no. 13E18CHA) and by the Graduate School CE within the Centre for Computational Engineering at the Technical University of Darmstadt.


\section*{References}
\bibliographystyle{ieeetr}
\bibliography{paperReferences}

\begin{thebibliography}{10}

\bibitem{teyber2020thermoeconomic}
R.~Teyber, L.~Brouwer, A.~Godeke, and S.~Prestemon, ``Thermoeconomic cost
  optimization of superconducting magnets for proton therapy gantries,'' {\em
  Superconductor Science and Technology}, vol.~33, no.~10, p.~105005, 2020.

\bibitem{lvovsky2013novel}
Y.~Lvovsky, E.~W. Stautner, and T.~Zhang, ``Novel technologies and
  configurations of superconducting magnets for {MRI},'' {\em Superconductor
  Science and Technology}, vol.~26, no.~9, p.~093001, 2013.

\bibitem{hfm_url}
Online, ``https://hfm.web.cern.ch/,'' consulted on April 2nd, 2024.

\bibitem{ravaioli2015cliq}
E.~Ravaioli, {\em {CLIQ}. A new quench protection technology for
  superconducting magnets}.
\newblock No.~CERN-THESIS-2015-091, Twente U., Enschede, 2015.

\bibitem{mulder2023external}
T.~Mulder, B.~Bordini, E.~Ravaioli, E.~Schnaubelt, M.~Wozniak, and A.~Verweij,
  ``External coil coupled loss induced quench ({E-CLIQ}) system for the
  protection of {LTS} magnets,'' {\em IEEE Transactions on Applied
  Superconductivity}, vol.~33, no.~5, pp.~1--5, 2023.

\bibitem{ravaioli2023optimizing}
E.~Ravaioli, T.~Mulder, A.~Verweij, and M.~Wozniak, ``Optimizing secondary
  {CLIQ} for protecting high-field accelerator magnets,'' {\em IEEE
  Transactions on Applied Superconductivity}, 2023.

\bibitem{carr1974ac}
W.~Carr~Jr, ``{AC} loss in a twisted filamentary superconducting wire. {I},''
  {\em Journal of Applied Physics}, vol.~45, no.~2, pp.~929--934, 1974.

\bibitem{wilson1983superconducting}
M.~N. Wilson, ``Superconducting magnets,'' 1983.

\bibitem{morgan1970theoretical}
G.~Morgan, ``Theoretical behavior of twisted multicore superconducting wire in
  a time-varying uniform magnetic field,'' {\em Journal of Applied Physics},
  vol.~41, no.~9, pp.~3673--3679, 1970.

\bibitem{turck1979coupling}
B.~Turck, ``Coupling losses in various outer normal layers surrounding the
  filament bundle of a superconducting composite,'' {\em Journal of Applied
  Physics}, vol.~50, no.~8, pp.~5397--5401, 1979.

\bibitem{ogasawara1980transient}
T.~Ogasawara, Y.~Takahashi, K.~Kanbara, Y.~Kubota, K.~Yasohama, and
  K.~Yasukochi, ``Transient field losses in multifilamentary composite
  conductors carrying {DC} transport currents,'' {\em Cryogenics}, vol.~20,
  no.~4, pp.~216--222, 1980.

\bibitem{campbell1982general}
A.~Campbell, ``A general treatment of losses in multifilamentary
  superconductors,'' {\em Cryogenics}, vol.~22, no.~1, pp.~3--16, 1982.

\bibitem{zhao20173d}
J.~Zhao, Y.~Li, and Y.~Gao, ``{3D} simulation of {AC} loss in a twisted
  multi-filamentary superconducting wire,'' {\em Cryogenics}, vol.~84,
  pp.~60--68, 2017.

\bibitem{Grilli2014}
F.~Grilli, E.~Pardo, A.~Stenvall, D.~N. Nguyen, W.~Yuan, and F.~Gomory,
  ``Computation of losses in {HTS} under the action of varying magnetic fields
  and currents,'' {\em {IEEE} Transactions on Applied Superconductivity},
  vol.~24, pp.~78--110, feb 2014.

\bibitem{escamez20163}
G.~Escamez, F.~Sirois, V.~Lahtinen, A.~Stenvall, A.~Badel, P.~Tixador,
  B.~Ramdane, G.~Meunier, R.~Perrin-Bit, and C.-E. Bruzek, ``3{D} numerical
  modeling of {AC} losses in multifilamentary {MgB}$_2$ wires,'' {\em IEEE
  Transactions on Applied Superconductivity}, vol.~26, no.~3, pp.~1--7, 2016.

\bibitem{riva2023h}
N.~Riva, A.~Halbach, M.~Lyly, C.~Messe, J.~Ruuskanen, and V.~Lahtinen,
  ``{H}-phi formulation in {S}parselizard combined with domain decomposition
  methods for modeling superconducting tapes, stacks, and twisted wires,'' {\em
  IEEE Transactions on Applied Superconductivity}, vol.~33, no.~5, pp.~1--5,
  2023.

\bibitem{olm2019simulation}
M.~Olm, S.~Badia, and A.~F. Mart{\'\i}n, ``Simulation of high temperature
  superconductors and experimental validation,'' {\em Computer Physics
  Communications}, vol.~237, pp.~154--167, 2019.

\bibitem{schnaubelt2024parallel}
E.~Schnaubelt, M.~Wozniak, J.~Dular, I.~C. Garcia, A.~Verweij, and S.~Schöps,
  ``Parallel-in-time integration of transient phenomena in no-insulation
  superconducting coils using {P}arareal,'' {\em ArXiv}, 2024.

\bibitem{nicolet2004modelling}
A.~Nicolet, F.~Zolla, and S.~Guenneau, ``Modelling of twisted optical
  waveguides with edge elements,'' {\em The European Physical Journal Applied
  Physics}, vol.~28, no.~2, pp.~153--157, 2004.

\bibitem{stenvall2012current}
A.~Stenvall, F.~Grilli, and M.~Lyly, ``Current-penetration patterns in twisted
  superconductors in self-field,'' {\em IEEE transactions on applied
  superconductivity}, vol.~23, no.~3, pp.~8200105--8200105, 2012.

\bibitem{dular2023helicoidal}
J.~Dular, F.~Henrotte, A.~Nicolet, M.~Wozniak, B.~Vanderheyden, and
  C.~Geuzaine, ``Helicoidal transformation method for finite element models of
  twisted superconductors,'' {\em in press in IEEE Transactions on Applied
  Superconductivity, doi:10.1109/TASC.2024.3416524}, 2024.

\bibitem{kameni2019reduced}
A.~Kameni, L.~Makong, F.~Bouillault, and P.~J. Masson, ``Reduced model to
  compute {AC} losses of twisted multifilamentary superconductors,'' {\em IEEE
  Transactions on Applied Superconductivity}, vol.~29, no.~7, pp.~1--6, 2019.

\bibitem{soldati2024ac}
L.~Soldati, M.~Breschi, P.~L. Ribani, T.~Spina, and C.-E. Bruzek, ``{AC} loss
  investigation in {MgB$_2$} multifilamentary wires: a numerical study.,'' {\em
  IEEE Transactions on Applied Superconductivity}, 2024.

\bibitem{breschi2008electromagnetic}
M.~Breschi and P.~L. Ribani, ``Electromagnetic modeling of the jacket in
  cable-in-conduit conductors,'' {\em IEEE transactions on applied
  superconductivity}, vol.~18, no.~1, pp.~18--28, 2008.

\bibitem{satiramatekul2010numerical}
T.~Satiramatekul and F.~Bouillault, ``Numerical modeling of superconducting
  filaments for coupled problem,'' {\em IEEE transactions on magnetics},
  vol.~46, no.~8, pp.~3229--3232, 2010.

\bibitem{satiramatekul2007magnetization}
T.~Satiramatekul, F.~Bouillault, A.~Devred, and D.~Leroy, ``Magnetization
  modeling of twisted superconducting filaments,'' {\em IEEE transactions on
  applied superconductivity}, vol.~17, no.~2, pp.~3737--3740, 2007.

\bibitem{satiramatekul2005contribution}
T.~Satiramatekul, {\em Contribution {\`a} la mod{\'e}lisation de l'aimantation
  des brins supraconducteurs}.
\newblock PhD thesis, Paris 11, 2005.

\bibitem{getdp}
P.~Dular, C.~Geuzaine, F.~Henrotte, and W.~Legros, ``A general environment for
  the treatment of discrete problems and its application to the finite element
  method,'' {\em IEEE Transactions on Magnetics}, vol.~34, no.~5,
  pp.~3395--3398, 1998.

\bibitem{vitrano2023open}
A.~Vitrano, M.~Wozniak, E.~Schnaubelt, T.~Mulder, E.~Ravaioli, and A.~Verweij,
  ``An open-source finite element quench simulation tool for superconducting
  magnets,'' {\em IEEE Transactions on Applied Superconductivity}, vol.~33,
  no.~5, pp.~1--6, 2023.

\bibitem{Bortot2017}
L.~Bortot, B.~Auchmann, I.~C. Garcia, A.~F. Navarro, M.~Maciejewski,
  M.~Mentink, M.~Prioli, E.~Ravaioli, S.~Schoeps, and A.~Verweij, ``{STEAM}: A
  hierarchical cosimulation framework for superconducting accelerator magnet
  circuits,'' {\em IEEE Transactions on applied superconductivity}, vol.~28,
  no.~3, pp.~1--6, 2017.

\bibitem{gmsh}
C.~Geuzaine and J.-F. Remacle, ``{G}msh: A 3{D} finite element mesh generator
  with built-in pre-and post-processing facilities,'' {\em International
  journal for numerical methods in engineering}, vol.~79, no.~11,
  pp.~1309--1331, 2009.

\bibitem{cati_steamAnalysis}
``{FiQuS} {CATI} strand,'' {\em
  https://gitlab.cern.ch/steam/analyses/cati-strand}, consulted on June 24th,
  2024.

\bibitem{verweij1997electrodynamics}
A.~P. Verweij, ``Electrodynamics of superconducting cables in accelerator
  magnets.,'' {\em Twente University, Enschede}, 1997.

\bibitem{goldacker2014roebel}
W.~Goldacker, F.~Grilli, E.~Pardo, A.~Kario, S.~I. Schlachter, and
  M.~Vojen{\v{c}}iak, ``Roebel cables from {REBCO} coated conductors: a
  one-century-old concept for the superconductivity of the future,'' {\em
  Superconductor Science and Technology}, vol.~27, no.~9, p.~093001, 2014.

\bibitem{jackson1999classical}
J.~D. Jackson, {\em Classical electrodynamics}.
\newblock AAPT, 1999.

\bibitem{dular2019finite}
J.~Dular, C.~Geuzaine, and B.~Vanderheyden, ``Finite-element formulations for
  systems with high-temperature superconductors,'' {\em IEEE Transactions on
  Applied Superconductivity}, vol.~30, no.~3, pp.~1--13, 2019.

\bibitem{ries1977ac}
G.~Ries, ``{AC}-losses in multifilamentary superconductors at technical
  frequencies,'' {\em IEEE Transactions on Magnetics}, vol.~13, no.~1,
  pp.~524--526, 1977.

\bibitem{geuzaine1999galerkin}
C.~Geuzaine, B.~Meys, F.~Henrotte, P.~Dular, and W.~Legros, ``A {G}alerkin
  projection method for mixed finite elements,'' {\em IEEE transactions on
  magnetics}, vol.~35, no.~3, pp.~1438--1441, 1999.

\bibitem{babuvska2003mixed}
I.~Babu{\v{s}}ka and G.~N. Gatica, ``On the mixed finite element method with
  {L}agrange multipliers,'' {\em Numerical Methods for Partial Differential
  Equations: An International Journal}, vol.~19, no.~2, pp.~192--210, 2003.

\bibitem{bossavit1988whitney}
A.~Bossavit, ``Whitney forms: A class of finite elements for three-dimensional
  computations in electromagnetism,'' {\em IEE Proceedings A-Physical Science,
  Measurement and Instrumentation, Management and Education-Reviews}, vol.~135,
  no.~8, pp.~493--500, 1988.

\bibitem{pellikka2013homology}
M.~Pellikka, S.~Suuriniemi, L.~Kettunen, and C.~Geuzaine, ``Homology and
  cohomology computation in finite element modeling,'' {\em SIAM Journal on
  Scientific Computing}, vol.~35, no.~5, pp.~B1195--B1214, 2013.

\bibitem{wilson2008nbti}
M.~N. Wilson, ``{NbTi} superconductors with low {AC} loss: A review,'' {\em
  Cryogenics}, vol.~48, no.~7-8, pp.~381--395, 2008.

\bibitem{rhyner1993magnetic}
J.~Rhyner, ``Magnetic properties and {AC}-losses of superconductors with power
  law current—voltage characteristics,'' {\em Physica C: Superconductivity},
  vol.~212, no.~3-4, pp.~292--300, 1993.

\bibitem{zachou2024}
G.~Zachou, M.~Wozniak, E.~Schnaubelt, T.~Mulder, J.~Dular, E.~Ravaioli, and
  A.~Verweij, ``A unified common source for material properties across
  simulation modelling tools,'' {\em https://cern.ch/smali}, 2024.

\bibitem{griffiths2010numerical}
D.~F. Griffiths and D.~J. Higham, {\em Numerical methods for ordinary
  differential equations: initial value problems}, vol.~5.
\newblock Springer, 2010.

\bibitem{shen2020review}
B.~Shen, F.~Grilli, and T.~Coombs, ``Review of the {AC} loss computation for
  {HTS} using {H} formulation,'' {\em Superconductor Science and Technology},
  vol.~33, no.~3, p.~033002, 2020.

\bibitem{bossavit2000remarks}
A.~Bossavit, ``Remarks about hysteresis in superconductivity modelling,'' {\em
  Physica B: Condensed Matter}, vol.~275, no.~1-3, pp.~142--149, 2000.

\bibitem{niyomsatian2020closed}
K.~Niyomsatian, J.~Gyselinck, and R.~V. Sabariego, ``Closed-form complex
  permeability expression for proximity-effect homogenisation of litz-wire
  windings,'' {\em IET Science, Measurement \& Technology}, vol.~14, no.~3,
  pp.~287--291, 2020.

\bibitem{klop2024electro}
C.~L. Klop, R.~Mellerud, C.~Hartmann, and J.~K. N{\o}land, ``Electro-thermal
  homogenization of {HTS} stacks and roebel cables for machine applications,''
  {\em IEEE Transactions on Applied Superconductivity}, 2024.

\bibitem{hopkins2023design}
S.~Hopkins, B.~Medina-Clavijo, C.~Barth, J.~Fleiter, and A.~Ballarino, ``Design
  optimization, cabling and stability of large-diameter high {$J_c$} {Nb$_3$Sn}
  wires,'' {\em IEEE Transactions on Applied Superconductivity}, vol.~33,
  no.~5, pp.~1--9, 2023.

\bibitem{schnaubelt2023electromagnetic}
E.~Schnaubelt, M.~Wozniak, S.~Sch{\"o}ps, and A.~Verweij, ``Electromagnetic
  simulation of no-insulation coils using {H}--phi thin shell approximation,''
  {\em IEEE Transactions on Applied Superconductivity}, vol.~33, no.~5,
  pp.~1--6, 2023.

\end{thebibliography}

\end{document}